\documentclass[english]{article}
\usepackage{filecontents}
\usepackage[T1]{fontenc}
\usepackage[latin9]{inputenc}
\usepackage{geometry}
\usepackage{mathrsfs}
\usepackage{amsmath}
\usepackage{amssymb}
\usepackage{setspace}

\makeatletter

\usepackage{mathrsfs}
\usepackage{babel}

\usepackage{babel}

\makeatother

\usepackage{babel}

\begin{document}
\title{\textbf{Emergence of Geometric phase shift in Planar Non-commutative
Quantum Mechanics}}
\author{{Saptarshi Biswas}\thanks{E-mail:saptarshibiswas531@gmail.com },\ {Partha
Nandi}\thanks{E-mail:parthanandi@bose.res.in }, \ {Biswajit Chakraborty}\thanks{E-mail:biswajit@bose.res.in }\\
 \\
\textit{{*} Indian Institute of Science Education and Research(IISER)}\\
\textit{Kolkata,Nadia 741246 West Bengal,India} \\
\textit{$^{\dagger,\ddagger}$ S.~N.~Bose National Centre for Basic
Sciences,} \\
\textit{JD Block, Sector III, Salt Lake, Kolkata-700106, India}}
\maketitle


\begin{abstract}
Appearance of adiabatic geometric phase shift in the context of non-commutative
quantum mechanics is studied using an exactly solvable model of 2D
simple harmonic oscilator (SHO) in Moyal plane, where momentum non-commutativity are
also considered along with spatial noncommutativity.
 For that we introduce a novel form of Bopp's shift, that bridges the non-commutative phase space operators with
their effective commutative counterparts, having their dependence on the non-commutative
parameters, and study the adiabatic evolution in the Heisenberg picture.
An explicit expression for the geometric phase shift under adiabatic
approximation is then found without using any perturbative technique. Lastly,
this phase is found to be related to the Hannay's angle of a classically
analogous system, by studying the evolution of the coherent state
of this system. 
\end{abstract}
\begin{large}

\section{Introduction}

Ever since M. Berry observed the occurrence of geometrical phase obtained
		in the adiabatic transport of a quantum system around a closed loop
		in the parameter space, the concept of Berry phase has attracted great
		interest both theoretically \cite{Berry_1985,PhysRevLett.56.893} as well as
		experimentally \cite{PhysRevLett.57.933,PhysRevLett.57.937}. Particularly it has
		been observed in certain condensed matter system,
		that this phase can give rise to effective non-commutative structure
		among the coordinates and thereby impacting physics
		\cite{PhysRevLett.95.137204,Duval:2005vn, BASU2010431}. Besides, the occurence of noncommutativity
		in the lowest Landau level in the Landau problem
		is quite well-known. Apart from these effective low energy effects, there
		are very strong plausibility arguments due to Doplicher et al.
		\cite{BF02104515,DOPLICHER199439}
		that these noncommutative algebra satisfied by spatio-temporal coordinates \cite{NANDI2017305}, when
		elevated to the level of operators, can naturally serve as a "deterrent"
		against gravitational collapse, associated with the localisation of
		an event at the Planck scale. Here the status of the noncommutaive parameters
		are more fundamental, as if they are new constants of Nature, like
		$\hbar,G,c$ etc \cite{Aschieri:2005zs} and possibly can play a
		vital role in the development of a future theory of quantum
		gravity\cite{Welling:1997qz}. This aspect was also corroborated in a seperate study of low energy
		limit of string theory by Seiberg-Witten \cite{Seiberg_1999}.\\
		
		Besides this above mentioned non-commutative structure among the space-time
		co-ordinates, it has also been proposed that along with the spatial
		components of space-time, momentum components too can satisfy a non-commutative
		algebraic structure \cite{PhysRev.72.874,Majid_1988,doi:10.1063/1.533331}. This was already
		indicated by  the reciprocity theorem proposed by Max Born way back in
		1938 \cite{doi:10.1098/rspa.1938.0060}. In the same spirit it was observed in \cite{PhysRevLett.111.101301,MADORE1992187} that the noncommutative structure among the spatial and momentum components can be related to the respective curvatures in momentum and coordinate  spaces respectively. Further, it was observed in
		\cite{ZHANG2004204} that, to maintain Bose-Einstein statistics in
		noncommutative spaces one needs to introduce noncommutative momenta as well. It was also indicated in the literature
			\cite{Duval:1993hs,bose1995,Horvathy:2010wv} that a nonrelativistic system in
			$(2+1)D$ admitting fractional spin can exhibit Galilean symmetry through a two-fold centrally extended
			Galilean algebra, where one involves the commutator of the boost generators $K_i$ between themselves, which is a
			non-vanishing constant $[K_1,K_2]=i\hslash\kappa$, and  other being the one involving boost and linear-momentum which gives the mass $m$:
			$[K_i,P_j]=i\hbar m\delta_{ij}$, with other commutators taking their usual
			forms. Here $\kappa$ can be associated with the fractional spin of the anyons, as has been shown in 
			\cite{PhysRevLett.49.957,Jackiw:2000tz}, through certain nonrelativistic reduction of (2+1)D Poincare algebra $iso(2,1)$. A minimalistic realization of this
			two-fold centrally extended algebra in terms of the covariantly transforming
			particle coordinate $\hat{x}_{i}$ under linear/angular momenta $(P_i,J)$ and
			boost $K_i$ for zero-values of the pair of Casimir operators, can only be
			provided if $\hat{x}_i$'s satisfy a NC algebra of the form $[\hat{x}_{i},
			\hat{x}_{j}]=i\theta \epsilon_{ij}$ with $\kappa=m^{2}\theta$ being a constant
			\cite{Horvathy:2002vt}. Further, if the planar system is now subjected to a
			static and uniform magnetic field B, then the momentum components too becomes
			noncommutative with further deformation in $[\hat{p}_1, \hat{p}_2]$ commutator
			\cite{Bellucci:2001xp,Horvathy:2002wc}. 
In fact the sheer presence of momentum noncommutativity alone can have nontrivial
		astrophysical consequences, like enhancing the  Chandrasekhar mass limit for the
		white dwarf stars, as has been shown by one of the authors very recently in
		\cite{Pal:2019awn}. The phase-space noncommutative structure has also been shown to
		emerge naturally in certain system in an enlarged phase space
		analysis \cite{Banerjee:2001zi}. It is therefore quite natural to investigate the
		occurrence
		of Berry phase, if any, in a quantum mechanical system where both
		position and momentum operators satisfy noncommutative algebra. This will then
		serve, in some sense, as the converse of the case, where the occurrence of Berry
		phase can give rise to noncommutative
		algebra \cite{PhysRevLett.95.137204}, as mentioned earlier. In order to
		undertake this study in this paper, we find it convenient to consider the
		simplest system of a harmonic oscilator lying in Moyal plane, where the momentum
		components are also taken to be noncommutative along with the position
		coordinates. Further, to compute Berry phase, we make the mass and "spring
		constant" to be time-dependent; varying adiabatically with time period $T$. There are precedence of
		such studies in the literature. For that one may
		cite the examples of well known Paul trap
		\cite{RevModPhys.62.531,PhysRevLett.66.527,PhysRevLett.67.3665} where  the "spring constant"/frequency is time
		dependent and in \cite{PhysRevA.20.1162,PhysRevA.32.1958} where the mass is also time
		dependent. In fact, our model (\ref{nch}) below was inspired by these previous works, although the Paul trap will not be directly applicable in our case, as it employs a varying magnetic field. Through a detailed analysis, we shall indeed show in this paper that
		occurrence of both types of non-commutativity
		in this quantum system plays a vital role for the
		existence
		of non-vanishing geometric phase shift for the system.\\

In this context, we would
			like to mention that some authors \cite{Bastos:2006kj} had also considered similar
			problem earlier, though 
			in a different system involving gravitational
			potential well and found no geometric contribution in the total adiabatic phase.
			However, there are certain basic differences between our considerations
			and their's. First of all, their original system, involving gravitational
			potential well, was commutative in nature to begin with and thereby possessed time reversal
			symmetry. On the other hand, it is known that the breaking of time reversal
			symmetry plays a necessary (but not sufficient) role in the existence of Berry's
			phase \cite{doi:10.1142/S0217979293002808,j.Ihm}. Also they introduced
			noncommutativity just by substituting the commutative variables by their NC
			counterparts, using the inverse of the Bopp shift (or Seiberg-Witten map in
			parlance of \cite{Bastos:2006kj}). Therefore, here the occurence of
			noncommutativity in the rewritten version, does not have any fundamental status;
			rather its occurence is a bit contrived in nature. More importantly, it
			has been pointed out in \cite{Bertolami:2005ud} that their particular form of non-commutative algebra and Bopp shift is
			equivalent to a scaled version of (\ref{ncms}) and its realization (\ref{bps1}) only for the scale factor $ \xi_{c}^{-1}$ as in (\ref{scale}).
			Consequently, the above mentioned  Bopp shift (\ref{bps1})
			becomes very restrictive in nature, in the sense that it provides a realization of the one
			parameter ($\xi$) family of phase space commutation relations (\ref{ncms}) in
			terms of commutative phase space variables (\ref{bps1}), only for a specific and
			critical value of $\xi=\xi_{c}$ (\ref{bd}). In contrast, our realization
			(\ref{bps3}) holds for any arbitrary value of $\xi$. And it is only for
			$\xi=\xi_{c}$ that these two realizations are unitarily equivalent. This has been
			elaborated in Appendix-A. In fact, as we show in the sequel, that the geometrical
			phase vanishes for that critical value of $\xi=\xi_{c}$, which can be shown by
			using any one of the realizations (\ref{bps1}) or (\ref{bps3}). And for other values of $\xi\ne\xi_{c}$, there is a non-vanishing geometrical phase.
			This can be clearly shown by making use of our realization (\ref{bps3}) only,
			which is solely responsible for the generation of a crucial dilatation term,
			whose presence, as we show below, is quite indispensable for getting the
			desired geometrical phase shift.\\
		
Furthermore, we would also like to point out that in the issue of Berry phase in noncommutative
			space was analyzed in \cite{Alavi_2003} also, albeit in absence of any momentum space noncommutativity;
			only position-position noncommutativity was considered. Besides, there the author
			basically computed the first order noncommutative correction, by expanding the
			Moyal star product upto $\mathcal{O}(\theta)$, to the already existing geometrical phase
			shift. This is in quite contrast to our analysis, where we show how the presence
			of phase-space noncommutativity itself can be a source for generating 
			non-vanishing Berry phase in an otherwise a simple quantum mechanical system, in the sense that this phase shift does not survive when the commutative limit is taken. To the best of our knowledge, this is the first such computation in this direction existing in literature. And this could be accomplished only through the novel form of the Bopp shift we have introduced, as mentioned above. Besides this, there  is yet another novelty in our analysis in the sense that, we have computed the geometric phase shift in Heisenberg picture. This served the dual purpose as it not only enabled us to virtually  read off the quantal Berry phase (we also showed how to obtain this by using the more well-known approch of obtaining the Berry phases through the one  acquired by the state vectors after time-evolving in Schrodinger picture) corresponding to our system-of-interest, just by looking at the extra phase-factor occurred over and above the dynamic phase by  adiabatically transporting the ladder operators of our system Hamiltonian, but also identify the corresponding Hennay angle in a rather straightforward manner. It therefore helps us to  "kill two birds with one stone".\\

		Moreover, an interrelation between the extra quantal geometric phase,  apart
		from the dynamical phase,  in the wave function in the quantum description  and
		the corresponding  angle shift at classical level was established by Berry
		through semiclassical torus quantization \cite{Berry_1985}. The change of the
		angle was found to be  related to the  rate of change of the extra phase with
		respect to the quantum number of the state which is being transported through
		adiabatically. This can be viewed as a manifestation of Bohr's correspondence
		principle for  phases arising through  adiabatic transports in the respective
		quantum and classical systems.
		However, in the present note,  we shall establish this classical correspondence
		by extending Berry's analysis to non-stationary   coherent states, in the spirit
		of \cite{Chaturvedi_1987,PhysRevLett.65.1697, PhysRevA.43.3217}, representing localized non-spreading wave packets 
		which are being transported along classical trajectories. \\
		\\
		The  paper is organised as follows: \\
		We begin by  introducing in sec-II a noncommutative
		phase space operators where we consider momentum space noncommutativity, along
		with the spatial ones of Moyal type. Here we also compute the instantaneous
		energy spectrum of two dimensional harmonic oscillator whose mass
		and frequency parameters are slowly varying with respect to time by
		making use of a generalised non-canonical phase space transformation, which we refer to as generalised Bopp shift (see Appendix-A), which maps the noncommutative
		phase
		space variables to their commutative counterparts. In sec-III we find
		out, in Heisenberg picture, the extra phase factor which  is  acquired by the creation and annihilation operators under an adiabatic excursion
		in parameter space of the system. We then discuss the geometric phase
		shift in state space of the oscillator in sec-IV and provide a contact
		between quantum geometric phase shift and classical Hannay angle 
		in sec V. Finally, we conclude the paper in sec-VI. Lastly , in the Appendix-A 
		we discuss different types of Bopp shifts i.e different realisations
		of non-commutative algebra and their relations with one another and in Appendix-B, apart from reviewing some of the necessary group theoretical aspects related to our model, we show that
		although the dilatation term in the Hamiltonian can apparently be transformed
		away by a time-dependent unitary transformation, it nevertheless re-appears in
		"disguise" in the dynamical term, albeit retaining its geometrical nature.

\section{Planar Non-Commutative system}

In this paper, we are basically considering a 2D harmonic oscillator on the Moyal plane, with time-dependent
coefficients $P(t)$ , $Q(t)$ varying adiabatically with period $T$:

\begin{equation}
\mathcal{H}(t)=P(t)(\hat{p}_{1}^{2}+\hat{p}_{2}^{2})+Q(t)(\hat{x}_{1}^{2}+\hat{x}_{2}^{2})\label{nch}
\end{equation}

such that $P(t),Q(t)>0$ and these time-dependent parameters are assumed to subsume
all other parameters like mass, frequency as mentioned in the previous section. This model is constructed in the spirit of \cite{RevModPhys.62.531,PhysRevLett.66.527,PhysRevLett.67.3665,PhysRevA.20.1162,PhysRevA.32.1958} where the plausibility of having time dependent parameters $P(t)$ and $Q(t)$ in real physical system was demonstrated. We are further assuming
that the momentum components also satisfy a non-commutative algebra
\cite{Ben_Geloun_2009,PhysRevD.90.024038,noncomm_formulations,rot_sym_non_commut}
in addition to the position coordinates, so that the entire non-commutative
structure takes the following form:

\begin{equation}
\begin{array}{c}
[\hat{x}_{i},\hat{x}_{j}]=i\theta\epsilon_{ij};[\hat{p}_{i},\hat{p}_{j}]=i\eta\epsilon_{ij};[\hat{x}_{i},\hat{p}_{j}]=i\hslash\delta_{ij}\end{array};\theta\eta<0\label{non-com}
\end{equation}\\
Note that we need to enforce $\theta\eta$ to be negative for consistent quantization. See for example \cite{Hatzinikitas:2001pm,Hatzinikitas:2001ht} and references therein.\\

\begin{doublespace}
In order to carry out the diagonalization, we introduce below a novel type of Bopp shift, providing a realisation of the above algebra (\ref{non-com}) through a linear map 
   $(\hat{x}_{i},\hat{p}_{j})\rightarrow(q_{i},p_{j});i,j=1,2$
, and we refer to this as generalized Bopp's shift (See Appendix-A for other kinds of Bopp shift  \cite{Bastos:2006kj,PhysRevLett.93.043002, Bertolami:2005ud}
and their relation to the one given bellow):

\begin{equation}
\begin{alignedat}{1}\hat{x}_{i} & =q_{i}-\frac{\theta}{2\hslash}\epsilon_{ij}p_{j}+\frac{\sqrt{-\theta\eta}}{2\hslash}\epsilon_{ij}q_{j}\\
\hat{p}_{i} & =p_{i}+\frac{\eta}{2\hslash}\epsilon_{ij}q_{j}+\frac{\sqrt{-\theta\eta}}{2\hslash}\epsilon_{ij}p_{j},
\end{alignedat}
\label{bpshift}
\end{equation}

\end{doublespace}

where $q_{i}$ and $p_{i}$ are commuting coordinates and momenta
respectively satisfying the usual Heisenberg algebra: $[q_{i},q_{j}]=0=[p_{i},p_{j}];[q_{i},p_{j}]=i\hslash\delta_{ij}$
and are distinguished by the absence of over head hats. Although
this transformation (\ref{bps2l}) is not a canonical one,
it nevertheless helps us in diagonalising the Hamiltonian. Substituting
(\ref{bpshift}) in (\ref{nch}) results in the following form of the Hamiltonian:

\begin{doublespace}
\begin{equation}
\mathcal{H}(t)=\alpha(t)(p_{1}^{2}+p_{2}^{2})+\beta(t)(q_{1}^{2}+q_{2}^{2})+\delta(t)(p_{i}q_{i}+q_{i}p_{i})-\gamma(t)(q_{1}p_{2}-q_{2}p_{1});\label{comh}
\end{equation}

\end{doublespace}

where the time-dependent coefficients $\alpha,\beta,\gamma,\delta$
are given by,

\begin{equation}
\begin{alignedat}{1}\alpha(t) & =P(t)\left\{ 1-\frac{\theta\eta}{4\hslash^{2}}\right\} +Q(t)\left(\frac{\theta}{2\hslash}\right)^{2}\\
\beta(t) & =Q(t)\left\{ 1-\frac{\theta\eta}{4\hslash^{2}}\right\} +P(t)\left(\frac{\eta}{2\hslash}\right)^{2}\\
\gamma(t) & =\frac{1}{\hslash}\left(\eta P(t)+\theta Q(t)\right)\\
\delta(t) & =\left(\frac{\sqrt{-\theta\eta}}{4\hslash^2}\right)\left(\eta P(t)-\theta Q(t)\right)
\end{alignedat}
\label{eq:alpha exp}
\end{equation}

At this stage we recognise the Hamiltonian as a combination of three
terms, 
\begin{equation}
\mathcal{H}(t)=\mathcal{H}_{gho,1}(t)+\mathcal{H}_{gho,2}(t)+\mathcal{H}_{\boldsymbol{L}}(t)
\label{hsep}
\end{equation}

where $\mathcal{H}_{gho,i}(t)'s$ \textit{(i=1 or 2)} are like generalised
time dependent harmonic oscillator Hamiltonian along $i^{th}-direction$,

\begin{equation}
\mathcal{H}_{gho,i}(t)=\alpha(t)(p_{i}^{2})+\beta(t)(q_{i}^{2})+\delta(t)(p_{i}q_{i}+q_{i}p_{i})\text{ }\text{ }\text{ }\text{ }\text{(no sum on \ensuremath{i})}\label{hghoi}
\end{equation}

and

\begin{equation}
\mathcal{H}_{\boldsymbol{L}}(t)=-\gamma(t)(q_{1}p_{2}-q_{2}p_{1})
\label{Zeem}
\end{equation}

which is like a Zeeman term. In order to diagonalize the whole Hamiltonian,
we first need to diagonalize $\mathcal{H}_{gho,i}(t)$ of (\ref{hghoi})
for each $i$ \cite{Xu_1996}\cite{Cervero_1989}\cite{theis_coherent_squeezed},
so that these Hamiltonians can be brought into the form $\mathcal{H}_{gho,i}(t)=X(t)(\boldsymbol{a_{i}^{\dagger}a_{i}}+1)$
\textit{(no sum on i)}. To that end we introduce annihilation (and corresponding creation)
operators $\boldsymbol{a}_{1},\boldsymbol{a}_{2}$
with the following structure:

\begin{equation}
\boldsymbol{a}_{j}=\left(\frac{\beta}{2\hslash\sqrt{\alpha\beta-\delta^{2}}}\right)^{1/2}\left[q_{j}+\left(\frac{\delta}{\beta}+i\frac{\sqrt{\alpha\beta-\delta^{2}}}{\beta}\right)p_{j}\right];\text{ }j=1,2\label{ol}
\end{equation}

satisfying $[\boldsymbol{a}_{i},\boldsymbol{a}_{j}^{\dagger}]=\delta_{ij}$.
Note that, we have $\beta>0$ and $\alpha\beta-\delta^{2}=\left(\frac{P\eta}{2\hslash}-\frac{Q\theta}{2\hslash}\right)^{2}+PQ>0$,
as follows from (\ref{eq:alpha exp}) and from the fact that $PQ>0$. The entire Hamiltonian (\ref{comh}) then takes the following form:

\begin{equation}
\mathcal{H}(t)=\hslash\omega\left(\sum_{j=1,2}\boldsymbol{a_{j}^{\dagger}a_{j}}+1\right)+i\hslash\gamma\left(\boldsymbol{a_{1}^{\dagger}a_{2}-a_{2}^{\dagger}a_{1}}\right);\text{ }\omega=2\sqrt{\alpha\beta-\delta^{2}}
\label{ttt}
\end{equation}

Noting at this stage \cite{PhysRevD.66.027701}, that the second non-diagonal
term, is like the Jordan-Schwinger representation of $J_{2}$ angular momentum
operator ($\overrightarrow{J}=\boldsymbol{a_{i}}^{\dagger}(\overrightarrow{\sigma})_{ij}\boldsymbol{a_{j}}$),
we need to carry out another additional unitary tranformation of the following form, which can
bring the term into the exact diagonal form of $J_{3}$, while retaining
the diagonal form of the first term:

\begin{equation}
\left[\begin{array}{c}
\boldsymbol{a_{1}}\\
\boldsymbol{a_{2}}
\end{array}\right]\text{ }\rightarrow\text{ }\left[\begin{array}{c}
\boldsymbol{a_{+}}\\
\boldsymbol{a_{-}}
\end{array}\right]=\frac{1}{\sqrt{2}}\left[\begin{array}{cc}
1 & -i\\
i & -1
\end{array}\right]\left[\begin{array}{c}
\boldsymbol{a_{1}}\\
\boldsymbol{a_{2}}
\end{array}\right]\label{eq:J2 to J3}
\end{equation}

\begin{equation}
\begin{array}{cc}
[\boldsymbol{a_{i},a_{j}^{\dagger}}]=\delta_{ij};[\boldsymbol{a_{i},a_{j}}]=0 & \text{ }(i,j\in\{+,-\})\end{array}
\end{equation}

Finally the diagonalized Hamiltonian in the standard quadratic form
reads,

\begin{equation}
\begin{alignedat}{1}\mathcal{H}(t) & =\hslash\sum_{j=+,-}\omega_{j}\boldsymbol{a_{j}}^{\dagger}\boldsymbol{a_{j}}+\hslash\omega;\text{ }\omega_{\pm}=\omega\mp\gamma\end{alignedat}
\label{dh}.
\end{equation}

Note that here we have identified two characteristic frequencies $\omega_{\pm}$
of the system.

 The eigenvalue equation of this Hamiltonian is,

\begin{equation}
\mathcal{H}(t)\left|n_{1},n_{2};t\right\rangle =E_{n_{1}n_{2}}(t)\left|n_{1},n_{2}; t\right\rangle ,
\end{equation}

whose  solution spectrum  can virtually be read-off from (\ref{dh}) as, 

\begin{equation}
\begin{array}{c}
E_{n_{1}n_{2}}(t)=\hslash\omega\left(n_{1}+n_{2}+1\right)-\hslash\gamma\left(n_{1}-n_{2}\right)\\
\left|n_{1},n_{2}; t \right\rangle =\frac{\left(\boldsymbol{a_{+}^{\dagger}}\right)^{n_{1}}\left(\boldsymbol{a_{-}^{\dagger}}\right)^{n_{2}}}{\sqrt{n_{1}!}\sqrt{n_{2}!}}\left|0,0;t\right\rangle 
\label{ene}
\end{array}
\end{equation}

where $n_{1},n_{2}$ are semipositive definite integers and $\boldsymbol{a}_{\boldsymbol{\pm}}(t)\left|0,0;t\right\rangle =0$. This reproduces the spectrum obtained in \cite{doi:10.1119/1.1624116}. Clearly the spectrum is non-degenerate and one can safely assume that there will not be any level crossing during the adiabatic process.
In this context, we would like to point out that essentially the same
system was analysed in \cite{Ben_Geloun_2009} but using Bartelomi's
realisation. One can easily check that both the algebra and spectrum
in \cite{Ben_Geloun_2009} agrees with (\ref{non-com}) and (\ref{ene}) respectively  by making the following
simple formal replacements in (\ref{ncms}) and (\ref{bps1}) : 
\begin{equation}
\theta\rightarrow\xi^{-1}\theta ;~ \eta\rightarrow\xi^{-1}\eta ;
~ \hbar\rightarrow\hbar_{eff}=\hbar\xi^{-1},
\label{scale}
\end{equation}
with $\xi=\xi_{c}$ (\ref{bd}).

Finally let us write down  our previous operators $\boldsymbol{a}_{1},\boldsymbol{a}_{2}$
of (\ref{ol}) in short as, 
\begin{equation}
\begin{alignedat}{1} & \boldsymbol{a}_{i}=A(t)(q_{i}+\left(B(t)+iC(t)\right)p_{i})\text{ ( 
	{i} \ensuremath{\in}\{1,2\})}\\
\text{where \text{ } } & A(t)=\left(\frac{\beta}{\hslash\omega}\right)^{1/2};~B(t)=\frac{\delta}{\beta};~C(t)=\frac{\omega}{2\beta}
\end{alignedat}
\label{eq:a in ABC}
\end{equation}

We see that $\boldsymbol{a_{\pm},a_{\pm}^{\dagger}}$ has explicit
time dependence through the time dependence of $A,B,C.$


\section{Heisenberg evolution of Ladder operators}

In this section we will solve the adiabatic evolution in Heisenberg
picture to look for the geometric phase shift. The  equation of motion for a generic operator $\hat {O} $ are given by  $\frac{d\hat{O}}{dt}=\frac{1}{i\hslash}[\hat{O},\mathcal{H}]+\frac{\partial\hat{O}}{\partial t}$. Specifically, for the ladder operators, they take the following forms:

\begin{equation}
\begin{array}{c}
\dfrac{d}{dt}\left[\begin{array}{c}
a_{+}\\
a_{-}\\
a_{+}^{\dagger}\\
a_{-}^{\dagger}
\end{array}\right]=\left[\begin{array}{cccc}
X_{+} & 0 & 0 & Y\\
0 & X_{-} & Y & 0\\
0 & Y^{*} & X_{+}^{*} & 0\\
Y^{*} & 0 & 0 & X^{*}
\end{array}\right]\left[\begin{array}{c}
a_{+}\\
a_{-}\\
a_{+}^{\dagger}\\
a_{-}^{\dagger}
\end{array}\right]\\
X_{\pm}=\frac{\dot{A}}{A}\pm i\left(\gamma\mp2C\beta\mp\frac{(\dot{B}+i\dot{C})}{2C}\right),Y=-\frac{(\dot{B}+i\dot{C})}{2C}
\end{array}\label{eq:ladder H evol}
\end{equation}

Uptill now all the  expressions we have found are exact. However, here
onwards we will start considering the adiabaticity of $P(t)$ and
$Q(t)$. Note, from the dependence of $A,B,C,\alpha,\beta,\gamma,\delta$
on $P(t),Q(t)$, we anticipate that they also follow the same order
of adiabaticity as $P$ and $Q$, i.e. if $\dot{P},\dot{Q}\approx\epsilon$
, $\ddot{P},\ddot{Q}\approx\epsilon^{2}\ldots$, then $\dot{F}\approx\epsilon$
, $\ddot{F}\approx\epsilon^{2}\ldots$ , where $F$ collectively stands
for $A,B,C,\alpha,\beta,\gamma,\delta$. We won't omit any term under adiabatic approximation right now, but
will only keep track of the order of various terms. Eventually, it
will be clear, that it is the second or higher order terms which are
ignorable \cite{book:17453,doi:10.1119/1.1285944}.

We can now decouple the four coupled equations occuring in (\ref{eq:ladder H evol})
by taking the derivatives of these equations and combining them suitably
to get:

\begin{equation}
\begin{array}{c}
\frac{d^{2}\boldsymbol{a}_{+}}{dt^{2}}=\frac{d\boldsymbol{a}_{+}}{dt}\left(X_{+}+\frac{\dot{Y}}{Y}+X_{-}^{*}\right)+\boldsymbol{a}_{+}\left(\dot{X}_{+}-\frac{\dot{Y}}{Y}X_{+}+YY^{*}-X_{+}X_{-}^{*}\right)\\
\frac{d^{2}\boldsymbol{a}_{-}}{dt^{2}}=\frac{d\boldsymbol{a}_{-}}{dt}\left(X_{-}+\frac{\dot{Y}}{Y}+X_{+}^{*}\right)+\boldsymbol{a}_{-}\left(\dot{X}_{-}-\frac{\dot{Y}}{Y}X_{-}+YY^{*}-X_{-}X_{+}^{*}\right)
\end{array}\label{eq:ladder 2nd diff}
\end{equation}

We can make some important observations here, if $Y\approx\epsilon$,
then $\dot{Y}\approx\epsilon^{2}$ , and so $\frac{\dot{Y}}{Y}\approx\epsilon$.

Now substituting $X_{+},X_{-},Y$ from (\ref{eq:ladder H evol}) and
only retaining terms involving $\frac{\dot{Y}}{Y}$, we deduce

\begin{equation}
\frac{d^{2}\boldsymbol{a}_{+}}{dt^{2}}=\frac{d\boldsymbol{a}_{+}}{dt}\left(\mathfrak{P}+\frac{\dot{Y}}{Y}\right)+\boldsymbol{a}_{+}\left(\mathfrak{Q}-\frac{\dot{Y}}{Y}X_{+}\right)\label{eq:ldev22}
\end{equation}\\
where,

\begin{equation}
\begin{aligned}\mathfrak{P}= & \left(2\frac{\dot{A}}{A}+2i\gamma+\frac{\dot{C}}{C}\right)\\
\mathfrak{Q}= & i\left(\dot{\gamma}-2\frac{d}{dt}\left(C\beta\right)-\gamma\frac{\dot{C}}{C}-2\frac{\dot{A}}{A}\gamma\right)+\left\{ \gamma^{2}-4C^{2}\beta^{2}-2\dot{B}\beta\right\} +\mathscr{O}\left(\epsilon^{2}\right)
\end{aligned}
\label{gopq}
\end{equation}

As one can check that the differential equation satisfied by $\boldsymbol{a}_{-}$
also has a similar form.

To proceed further we now need to cast (\ref{eq:ldev22}) into its,
the so-called, normal form. To that end, we define another time-dependent
operator \textbf{$\boldsymbol{b}(t)$ }as,

\begin{equation}
\boldsymbol{a_{+}}(t)=\boldsymbol{b}(t)e^{\frac{1}{2}\int_{t}\left(\mathfrak{P}+\frac{\dot{Y}}{Y}\right)d\tau},\label{eq:normaling}
\end{equation}

 In terms of $\boldsymbol{b}(t) $ the equation (\ref{eq:ldev22}) can now be re-written as ,

\begin{equation}
\boldsymbol{\ddot{b}}+\boldsymbol{b}\left(\frac{\dot{\mathfrak{P}}}{2}-\frac{\mathfrak{P}^{2}}{4}-\mathfrak{Q}\right)+\boldsymbol{b}\left(\frac{\dot{Y}}{Y}X_{+}-\frac{\mathfrak{P}}{2}\frac{\dot{Y}}{Y}+\mathscr{O}\left(\epsilon^{2}\right)\right)=0\label{beq}
\end{equation}

Now, we have $\frac{\dot{Y}}{Y}=\frac{\ddot{B}+i\ddot{C}}{\dot{B}+i\dot{C}}-\frac{\dot{C}}{C}$ as follows from (\ref{eq:ladder H evol}).
Let's write it as

\begin{equation}
\frac{\dot{Y}}{Y}=Z+i\tilde{Z}-\frac{\dot{C}}{C};
\end{equation}

here both $Z$ and $\tilde{Z}$$\approx\mathscr{O}\left(\epsilon\right)$
and corresponds respectively to the real and imaginary part of $\frac{\ddot{B}+i\ddot{C}}{\dot{B}+i\dot{C}}$.

Then using the expressions of $\mathfrak{P}$ and $\mathfrak{Q}$
from (\ref{gopq}) we get,

\begin{equation}
\boldsymbol{\ddot{b}}+\boldsymbol{b}(U+iV)=0
\end{equation}

where,

\begin{equation}
\begin{alignedat}{1}U= & 4C^{2}\beta^{2}+2\dot{B}\beta+2\tilde{Z}C\beta+\mathscr{O}\left(\epsilon^{2}\right)\approx\mathscr{O}\left(\epsilon^{0}\right)\\
V= & 2\frac{d}{dt}\left(C\beta\right)-2C\beta \left(Z-\frac{\dot{C}}{C}\right)+\mathscr{O}\left(\epsilon^{2}\right)\approx\mathscr{O}\left(\epsilon\right)
\end{alignedat}
\end{equation}

Note that, since we are working in adiabatic regime, the functions
$U$ and $V$ vary very very slowly with time. Hence, the formula
for WKB approximation for complex potential\cite{article} can be
applied to get the general solution of the differential equation as,

\begin{normalsize}
\begin{equation}
\boldsymbol{b}(t)=\boldsymbol{b}(0)\left[\frac{C_{1}}{\sqrt{\left|\xi(t)\right|}}\exp\left(\int_{0}^{t}\left(i\xi(\tau)-\phi(\tau)\right)d\tau\right)+\frac{C_{2}}{\sqrt{\left|\xi(t)\right|}}\exp\left(\int_{0}^{t}\left(-i\xi(\tau)+\phi(\tau)\right)d\tau\right)\right]\label{eq:cwkbsol}
\end{equation}
\end{normalsize}
where, $\sqrt{U+iV}=\xi+i\phi$ and ($C_{1}$, $C_{2}$) are arbitrary coefficients. This result can also be derived by
solving the differential equation using the method of successive approximation
and considering the adiabatic variation of U and V.

In our case, this boils down to 
\begin{normalsize}
\begin{equation}
\begin{alignedat}{1}\xi=\sqrt{\frac{\sqrt{U^{2}+V^{2}}+U}{2}} & \approx\sqrt{U+\frac{V^{2}}{4U}}\approx\sqrt{U}\approx2C\beta+\frac{\dot{B}\beta+C\beta\tilde{Z}}{2C\beta}\\
\phi=\sqrt{\frac{\sqrt{U^{2}+V^{2}}-U}{2}} & \approx\sqrt{\frac{V^{2}}{4U}}\approx\frac{2\frac{d}{dt}\left(C\beta\right)-2C\beta\left(Z-\frac{\dot{C}}{C}\right)}{4C\beta}
\end{alignedat}
\end{equation}
\end{normalsize}

Note that we have ignored second and higher order terms. We now observe
that the solution must satisfy the boundary condition: $\boldsymbol{b}(t=0)=\boldsymbol{b}(0)$
. Also, the periodicity of the parameters imply $\sqrt{\left|\xi(0)\right|}=\sqrt{\left|\xi(T)\right|}$.
Finally, it can be observed that, only the second term with coefficient $C_{2}$ in the solution
(\ref{eq:cwkbsol}), yields the dynamical phase of $\boldsymbol{a}_{+}$
with proper sign. This will eventually be clear as we calculate $\boldsymbol{a}_{+}(T).$ We therefore set $ C_{1}=0$ in (\ref{eq:cwkbsol}). Now combining all these expressions, the particular solution of
(\ref{beq}) is obtained as:

\begin{equation}
\begin{aligned}\boldsymbol{b}(T) & \approx\boldsymbol{b}(0)\exp\left(\int_{0}^{T}\left\{ -i\left(2C\beta+\frac{\dot{B}\beta+C\beta\tilde{Z}}{2C\beta}\right)+\phi\right\} d\tau\right)\end{aligned}
\end{equation}

Now, we had, $\frac{\dot{Y}}{Y}=Z+i\tilde{Z}-\frac{\dot{C}}{C}$.
As the latter is an exact
differential, we can write, 
\begin{equation}
\int_{0}^{T}\frac{\dot{Y}}{Y}d\tau=\int_{0}^{T}\left(Z+i\tilde{Z}-\frac{\dot{C}}{C}\right)d\tau=\int_{0}^{T}Zd\tau+i\int_{0}^{T}\tilde{Z}d\tau=0
\end{equation}

So, $\int_{0}^{T}Zd\tau=\int_{0}^{T}\tilde{Z}d\tau=0$ , implying
that, $\phi$ is also an exact differential.

Hence using (\ref{eq:normaling}), we can essentially drop the term
involving just the exact derivatives and then split the respective
dynamical and geometric phase shifts as, 

\begin{normalsize}
\begin{equation}
\begin{alignedat}{1}\boldsymbol{a}_{+}(T) & =\boldsymbol{a}_{+}(0)\exp\left(-i\int_{0}^{T}\left(2C\beta+\frac{\dot{B}\beta+C\beta\tilde{Z}}{2C\beta}\right)d\tau+\frac{1}{2}\int_{0}^{T}\left(\frac{\dot{A}}{A}+2i\gamma+\frac{\dot{C}}{C}+\frac{\dot{Y}}{Y}\right)d\tau\right)\\
 & =\boldsymbol{a}_{+}(0)\exp\left(-i\int_{0}^{T}\left[\left(2C\beta-\gamma\right)+\left(\frac{\dot{B}\beta+C\beta\tilde{Z}}{2C\beta}\right)\right]d\tau\right)
\end{alignedat}
\end{equation}
\end{normalsize}

And the solution becomes, 
\begin{equation}
\begin{aligned}
\boldsymbol{a}_{+}(T) & =\boldsymbol{a}_{+}(0)\exp\left(-i\int_{0}^{T}\left[\left(2C\beta-\gamma\right)+\left(\frac{\dot{B}}{2C}\right)\right]d\tau\right)\\
 & =\boldsymbol{a}_{+}(0)\exp\left(-\frac{i}{\hslash}\int_{0}^{T}\left(\hslash\omega-\gamma\hslash\right)d\tau-i\int_{0}^{T}\frac{\beta}{\omega}\frac{d}{d\tau}\left(\frac{\delta}{\beta}\right)d\tau\right),\label{eq:a+sol}
\end{aligned}
\end{equation}

with the two terms in the exponent representing the dynamical and
the geometrical phases, respectively.

Finally, a close look into the decoupled evolution equation of $\boldsymbol{a}_{-}$
in (\ref{eq:ladder 2nd diff}) tell us that, it is similar to the
one for $\boldsymbol{a}_{+}$, except that $(+\gamma)$ is replaced
by $(-\gamma)$. Also, $\gamma$ is entering into the solution only
through the substitution of (\ref{eq:normaling}). So, we get

\begin{equation}
\begin{aligned}
\boldsymbol{a}_{-}(T) & =\boldsymbol{a}_{-}(0)\exp\left(-\frac{i}{\hslash}\int_{0}^{T}\left(\hslash\omega+\gamma\hslash\right)d\tau-i\int_{0}^{T}\frac{\beta}{\omega}\frac{d}{d\tau}\left(\frac{\delta}{\beta}\right)d\tau\right),\label{eq:a-sol}
\end{aligned}
\end{equation}

which gives the correct dynamical phase for $\boldsymbol{a}_{-}$
.


\section{Geometric phases}
Now looking  at the  second phase factor  in the expression of both the creation and annihilation operators $\boldsymbol{a}_{\pm}(T)  $ in (\ref{eq:a+sol}) and (\ref{eq:a-sol}), the additional  factor over and above  the dynamical phase,   obtained  by  leading behaviour for adiabatic transport around a closed loop $ \Gamma$  in time $ T $  can be identified with the  Berry phase or geometric phase ( more precisely geometric phase shift ) in the Heisenberg picture. As pointed out earlier, the result obtained  here can readily be converted to the more familiar form in terms of the phase gathered by the state vector by going over from the Heisenberg to the Schrodinger picture.
 The geometric phase shift $\Phi_{G}$ found above can be written in
a more familiar form by using the transformation $\frac{d}{d\tau}=\frac{d\boldsymbol{R}}{d\tau}\cdot\nabla_{\boldsymbol{R}}$
, where \textbf{$\boldsymbol{R}$} represents a vector in the parameter-space
whose components are the time dependent. Then
we can write $\Phi_{G}$ as a line-integral over
a closed loop $\Gamma$ traced out in the parameter space as $\tau$
varies from 0 to $T$ , i.e. a complete period so that it  can be written as a functional of $\Gamma$ as,

\begin{equation}
\Phi_{G}[\Gamma]=\oint_{\Gamma=\partial S}\frac{\beta}{\omega}\nabla_{\boldsymbol{R}}\left(\frac{\delta}{\beta}\right)\cdot d\boldsymbol{R}=\iint_{S}\nabla_{\boldsymbol{R}}\left(\frac{\beta}{\omega}\right)\times\nabla_{\boldsymbol{R}}\left(\frac{\delta}{\beta}\right)\cdot d\boldsymbol{S}\label{eq:ladderphase},
\end{equation}

where we have made use of Stoke's theorem in the second equality, to recast it as a surface integral over $S$.
Note that $ S $ stands for any surface belonging to the equivalence class of surfaces in the parameter space having the same boundary $ \Gamma $ , and where any two such  surfaces can be connected by smooth deformation without encountering any singularity. Now  substituting $\alpha,\beta,\gamma,\delta$ from (\ref{eq:alpha exp}),
the geometric phase can now be expressed in terms of our original
time dependent parameters $P(t)$, $Q(t)$ and the  noncommutative  parameters  $\theta$ and $\eta$ as,

\begin{equation}
\begin{aligned}\Phi_{G}[\Gamma]= & \left( \frac{\sqrt{-\theta \eta}}{4\hbar}\right) \iint_{ S}\nabla_{\boldsymbol{R}}\left(\frac{Q\left(1-\frac{\theta\eta}{4\hslash^2}\right)+P\left(\frac{\eta}{2\hslash}\right)^{2}}{\sqrt{\left(P\left(\frac{\eta}{2\hslash}\right)-Q\left(\frac{\theta}{2\hslash}\right)\right)^{2}+PQ}}\right)\times\\
 & \nabla_{\boldsymbol{R}}\left(\frac{P\left(\frac{\eta}{2\hslash}\right)-Q\left(\frac{\theta}{2\hslash}\right)}{Q\left(1-\frac{\theta\eta}{4\hslash^2}\right)+P\left(\frac{\eta}{2\hslash}\right)^{2}}\right)\cdot d\boldsymbol{S}
\end{aligned}
\label{gphase in noncommute}
\end{equation}

One can be rest assured at this stage, that the denominator never
vanishes as PQ>0. Also it is worth noting that, in the absence of
either of the two types of non-commutativity i.e. if $\theta$ or $\eta=0$,
the geometric phase vanishes. So, it is the non-commutative nature
of phase space, as a whole, alongside the geometry of the parameter space trajectory,
which plays crucial role on the appearance of geometric phase shift for
this 2D harmonic oscillator system.\\

Before we proceed further, let us pause for a while and make some pertinent comments:

\begin{itemize}	
	
\item	The reason behind the appearance of this phase can be attributed, in our case, to the time reversal symmetry breaking of the Hamiltonian  (\ref{nch},\ref{comh})  \cite{doi:10.1142/S0217979293002808,j.Ihm,book:15453,Time,U}. To elaborate on this matter, we need to explain in bit detail about what we mean by time reversal symmetry of a generic  time-dependent Hamiltonian $\mathcal{H}(t) $ having a set of time dependent parameters. First note that this Hamiltonian $\mathcal{H}(t)$ can be regarded as a sequence of instantaneous time-independent Hamiltonians: one for each time $t$ and each of them being a distinct Hermitian matrix (finite or infinite) with real diagonal and complex off-diagonal entries, in general. Now, time reversal symmetry refers to the instantaneous Hamiltonians $ \mathcal{H}(t_{0})$, i.e. as if the parameters are frozen at their values corresponding to that instant $t=t_{0}$, which is not time-dependent anymore. And for a time-dependent Hamiltonian being time-reversal symmetric means that, each such instantaneous Hamiltonians in the sequence must be real symmetric, not just complex Hermitian. In other words, if we let a system evolve by this Hamiltonian $\mathcal{H}(t_0)$ for some
finite time interval after $t_0$ and then time reverse at any later time $t > t_0 $, then the
corresponding wave function is simply obtained just by complex conjugation, without touching  the set of parameters occurring there at all. This is because of the fact that the values of the parameters are now held fixed to their respective values at time $t_0$ and so are affected neither by the subsequent continuous time evolution nor under the discrete time-reversal transformation i.e. the time arguments occurring in the parameters undergo no flipping of sign  by this, which we can call more appropriately a `` quasi-time
reversal" transformation. Consequently, under this
`` quasi-time reversal" transformation, the system will retrace its own history and if that happens  regardless of which instantaneous Hamiltonian of the original time-dependent system was chosen, then we say $\mathcal{H}(t) $ is time-reversal symmetric. A concrete mathematical definition of such time-reversal symmetry \cite{doi:10.1142/S0217979293002808,PhysRevLett.67.251} would be:
\begin{equation*}
\begin{aligned}
\hat{\Xi}~\hat{H}(t)~\hat{\Xi}^{-1} &= \hat{H}(t)  &&  \text{(without any change in the sign of $t$)}
\end{aligned}
\end{equation*}
where, the antilinear (quasi or instantaneous) time-reversal operator $\hat{\Xi}$  leaves all the real time-dependent parameter(s) intact. The latter nomenclature i.e. instantaneous time-reversal symmetry has been borrowed  from \cite{book:174801},
	where  similar circumstances was encountered in a system involving topological insulator.

 Now, in commutative plane: $\mathcal{H}_{c}(t)=P(t)(\hat{p}_{1}^{2}+\hat{p}_{2}^{2})+Q(t)(\hat{x}_{1}^{2}+\hat{x}_{2}^{2})$;
with $\hat{p}_{1},\hat{p}_{2},\hat{x}_{1},\hat{x}_{2}$ satisfying
ordinary Heisenberg algebra. (Instantaneous or quasi) time reversal transformation operates as: $\hat{p}_{i}\rightarrow\hat{p}'_{i}=\hat{\Xi}~\hat{p}_{i}~\hat{\Xi}^{-1}=-\hat{p}_{i}$
and $\hat{x}_{i}\rightarrow\hat{x}'_{i}=\hat{\Xi}~ \hat{x}_{i}~\hat{\Xi}^{-1}=\hat{x}_{i}$
, shows that the Hamiltonian is symmetric under time reversal: $\hat{\Xi}~\mathcal{H}_{c}(t)~\hat{\Xi}^{-1}=\mathcal{H}_{c}(t)$, as the parameters $P(t)$ and $Q(t)$ are not touched.\\

On the other hand, in the non-commutative plane, the dynamics is given by the Hamiltonian (\ref{nch}) with non-commutative coordinates and momenta satisfying algebra (\ref{ncms}) or equivalently by the Hamiltonian
(\ref{comh}) with the mathematically commuting coordinates and momentum,
transforming like, $p_{i}\rightarrow-p_{i}$ , $q_{i}\rightarrow q_{i}$,
under time reversal. Hence, the Hamiltonian $\mathcal{H}(t)=\alpha(t)(p_{1}^{2}+p_{2}^{2})+\beta(t)(q_{1}^{2}+q_{2}^{2})+\delta(t)(p_{i}q_{i}+q_{i}p_{i})-\gamma(t)(q_{1}p_{2}-q_{2}p_{1})$
is not time reversal symmetric: $\hat{\Xi}~\mathcal{H}(t)~\hat{\Xi}^{-1}\neq\mathcal{H}(t)$;
 the presence of the dilatation term and the Zeeman like
term breaks this symmetry\cite{articles}. Particularly, the breaking by dilatation term is primarily
responsible for getting nonvanishing Berry phase in our case. In fact it
has been shown in \cite{doi:10.1142/S0217979293002808,PhysRevLett.67.251}
that this time reversal symmetry breaking is a necessary, but not sufficient,
condition for the appearance of non-vanishing Berry phase. And it
is because of this broken time reversal symmetry, while considering noncommutative phase space \cite{PhysRevLett.67.251,PhysRevA.46.5358,Ben_Geloun_2009}, that there arises the possibility of obtaining a non-vanishing geometrical phase shift in our system of 2D SHO in noncommutative phase space.\\

	\item The Berry connection one-form $\boldsymbol{A}$  on the loop $\Gamma $ that we found can be also be directly read-off from (\ref{eq:a+sol}) and (\ref{eq:a-sol}) as:

	\begin{equation}
	 \boldsymbol{ A}=\frac{\beta}{\omega}d(\frac{\delta}{\beta}) =-\frac{\alpha}{\omega}d(\frac{\delta}{\alpha})-d[tan^{-1}(\sqrt{\frac{\alpha \beta}{\delta^{2}} -1})],
	 \label{phg}
  \end{equation}
  
  showing that, upto a non-singular gauge transformation, the Berry connection can also be written as 
  \begin{equation}
  \boldsymbol{ A}:=-\frac{\alpha}{\omega}d(\frac{\delta}{\alpha}) 
  \label{connec}
  \end{equation}
This particular feature of this connection one-form is indeed quite gratifying as the symmetry between $\alpha$ and $\beta$, the coefficients of $\vec{p}^{2}$	 and $\vec{q}^{2}$ in the Hamiltonian (\ref{comh}), is some how restored with this. In fact, this form of the connection one form (\ref{connec}) occurred earlier in \cite{Berry_1985,Hannay_1985,Giavarini:1988wy,doi:10.1142/S0217751X94001047}, where a Hamiltonian of the same form as (\ref{comh}) was used to describe 1D parametric generalized harmonic oscillator, except that there time-dependent coefficients  $\alpha, \beta, \delta$ were of fundamental nature by themselves, unlike in our case, where $\alpha, \beta, \delta$ are not fundamental and rather given in terms of more fundamental $P$ and $Q$ through a set of linear relations (\ref{eq:alpha exp}). Also note that any of our time-dependent parameters $\alpha(t), \beta(t), \delta(t), \gamma(t)$ cannot vanish for all time $t$, otherwise we would have got $P(t)\propto Q(t)$\footnote{For example, $\gamma(t)=0~~\forall t$, implies from (\ref{eq:alpha exp}), that $P(t)\propto Q(t) ~~\forall t$. In fact writing more explicitly, we have $\frac{P}{Q}=-\frac{\theta}{\eta}$. } as is clear from (\ref{eq:alpha exp}). With this the closed loop $\Gamma$ will collapse to a 1D line in $P,Q$ space thereby yielding a vanishing geometric phase. This is particularly true for $\delta(t)$ which clearly plays a vital role here and its origin can be traced back to (\ref{ol}) where ($\frac{\delta}{\beta}$) occurs as the real part of the coefficient $p_j$; in its absence we cannot get any geometrical phase, as is clear from (\ref{eq:a+sol},\ref{eq:a-sol}). \\

\item 
	
Although the $\gamma (t)$ occurring in the Zeeman term in (\ref{comh}) is required to be
	non-vanishing, in order to get a non-vanishing Berry phase, it also plays another important role by allowing us to avoid the crossings of energy levels by lifting the degeneracy, as we have mentioned already in Section 2. Despite all these, it
	does not have its explicit presence in the expression of the Berry phase (\ref{eq:ladderphase},\ref{connec})(if we ignore the relations in (\ref{eq:alpha exp}) for the time being); it
	rather appears in the dynamical phases in (\ref{eq:a+sol},\ref{eq:a-sol}) and corroborates the general observation made by J. Anandan et al.\cite{PhysRevD.35.2597} where the dynamical group was $ U(2)$. To understand the reason behind
all this, observe that, although the Hamiltonian $\mathcal{H} (t)$ (\ref{comh},\ref{hsep}) do not commute at different times: $[\mathcal{H}(t), \mathcal{H}(t')]\ne 0$ for $t \ne t'$, but being an element of the algebra $ su(1,1)\oplus u(1) $ it splits into two commuting parts as in (80) (see Appendix B). Importantly, these two terms in (80) commute with each other at different time also. Consequently, the corresponding time evolution operator
(in Schrodinger picture) factorizes as,
\begin{equation}
\mathcal{U}=\hat{\mathcal{T}}(e^{-\frac{i}{\hslash}\int dt[\alpha(t)(p_{1}^{2}+p_{2}^{2})+\beta(t)(q_{1}^{2}+q_{2}^{2})+\delta(t)(p_{i}q_{i}+q_{i}p_{i})]})\hat{\mathcal{T}}(e^{\frac{i}{\hslash}\int dt\gamma(t)(q_{1}p_{2}-q_{2}p_{1})}).
\label{angul}
\end{equation}
where $\hat{\mathcal{T}}$ is the time-odering or chronological operator.
After all, it can be easily seen  that
$\gamma (t)$, like $\omega (t)$, occurs in the integral as $\int_0^T \gamma (t) dt $,
which is not a functional of the closed loop $\Gamma$, the latter being the
telltale sign of geometrical phases (\ref{eq:ladderphase},\ref{phg}): $\Phi_G[\Gamma]=\int_{\Gamma} \boldsymbol{A}$.

\item 
	
Finally, we would like to study the implication of the unitary equivalence
			between the two forms of Bopp-shifts (\ref{bps1},\ref{bps3}), as has been
			demonstrated in Appendix A, in our context. To begin with, note that the
			geometrical phase $\Phi_{G}$ in (\ref{gphase in noncommute}) was obtained for
			the scale parameter $\xi$, taken without loss of generality, to be  $\xi=1$. For
			any other value of $\xi$,  the corresponding $\Phi_{G}$ can be obtained easily
			by replacing $\theta\rightarrow\xi\theta$  and $\eta\rightarrow\xi\eta$, i.e. make use of the Bopp shift (\ref{bps3}). Note
			that $\Phi_{G}$ remains invariant iff $\hbar$ is also scaled to
			$\hbar\rightarrow\xi\hbar$, as (\ref{bps3}) will reduce in this case to $\xi=1$, thereby undoing the scaling operation just as in (\ref{scale}). At the critical point $\xi=\xi_{c}$ (\ref{bd}), however, the counterpart of the expression (\ref{gphase in noncommute}) will be absent if one makes use of realisation (\ref{bps1}). On the other hand, with the equivalent realisation (\ref{bps3}) with $ \xi=\xi_{c}$, the counterpart of (\ref{gphase in noncommute}) will definitely be present, but it can be shown that it loses its geometrical nature and in either case, one finds that $\Phi_{G}$ vanishes: $\Phi_{G}=0$. This can be shown in two different but equivalent ways by taking advantage of the
			above-mentioned unitary equivalence between two forms of Bopp shifts (\ref{bps1},\ref{bps3}) holding for $ \xi=\xi_{c}$ (\ref{bd}) only. To this end, we first employ (\ref{bps1}) in (\ref{nch}) to obtain the
			Hamiltonian in the following form,

\begin{equation}
			\mathcal{H}^{1}(t)=\alpha^{(1)}(t)p^{2}_{i}+\beta^{(1)}(t)q^{2}_{i}-\gamma^{(1)}(t)\epsilon_{ij}q_{i}p_{j}
			\label{nh}
\end{equation}
where,
\begin{equation}
\begin{alignedat}{1}\alpha^{(1)}(t) &
			=\xi_{c}\left(P(t)+\frac{\theta^{2}}{4\hbar^{2}}Q(t)\right)\\
			\beta^{(1)}(t) & =\xi_{c}\left(Q(t)+\frac{\eta^{2}}{4\hbar^{2}}P(t)\right)\\
			\gamma^{(1)}(t) & =\frac{\xi_{c}}{\hbar}\left(\theta Q(t)+\eta P(t)\right)
\end{alignedat}
			\label{eq:alpha exp2}
\end{equation}\\
This pair (\ref{nh},\ref{eq:alpha exp2}) of equations,  are the counterparts of
				(\ref{comh}) obtained by employing (\ref{bpshift}) in
				(\ref{nch}). Here, not only we have $\delta^{1}(t)=0 $ $\forall t$ (which is the
				counterpart of $\delta(t)$ in (\ref{eq:alpha exp})), but also note the absence
				of any linear equation relating $\delta^{(1)}(t)$ with $P(t)$ and $Q(t)$, thereby indicating the absence of a counterpart of (\ref{gphase in noncommute}) following from (\ref{eq:ladderphase}), in this case. Consequently, one has to make use of (\ref{phg},\ref{connec}), or the second terms in the
				exponents of (\ref{eq:a+sol}) and (\ref{eq:a-sol}), to find that Berry phase vanishes: $ \Phi_{G}=0$, in this case.
	
	To arrive at the same conclusion through the Bopp Shift (\ref{bps3}), we need
			to write down the corresponding expressions for
			$\alpha$,$\beta$,$\gamma$,$\delta$ for $\xi=\xi_{c}$ which is simply obtained by
			replacing $\theta\rightarrow\xi_{c}\theta;\eta\rightarrow\xi_{c}\eta$ in
			(\ref{eq:alpha exp}), as mentioned above, to get
	 
\begin{equation}
\mathcal{H}^{(2)}(t)=\alpha^{(2)}(t)p^{2}_{i}+\beta^{(2)}(t)q^{2}_{i}-\gamma^{(2)}(t)\epsilon_{ij}q_{i}p_{j}+\delta^{(2)}(t)(q_{i}p_{i}+p_{i}q_{i}) 
\label{dc}
\end{equation}
	 
 where

\begin{equation}
\begin{alignedat}{1}\alpha^{(2)}(t)=\alpha(t;\xi_{c}) & =P(t)\left\{ 1-\frac{\xi_{c}^{2}
	\theta\eta}{4\hslash^{2}}\right\} +Q(t)\left(\frac{\xi_{c}\theta}{2\hslash}\right)^{2}\\
\beta^{(2)}(t)=\beta(t;\xi_{c}) & =Q(t)\left\{ 1-\frac{\xi^{2}_{c}\theta\eta}{4\hslash^{2}}\right\} +P(t)\left(\frac{\xi_{c}\eta}{2\hslash}\right)^{2}\\
\gamma^{(2)}(t)=\gamma(t;\xi_{c}) & =\frac{\xi_{c}}{\hbar}\left(\eta P(t)+\theta Q(t)\right)\\
\delta^{(2)}(t)=\delta(t;\xi_{c}) & =\left(\frac{\xi^{2}_{c}\sqrt{-\theta\eta}}{4\hslash^2}\right)\left(\eta P(t)-\theta Q(t)\right)
\end{alignedat}
\label{eq:alpha exp3} 
\end{equation} \\

Apparently the presence and non-vanishing nature of $\delta^{(2)}(t)$ here, however, suggests a non-vanishing
$\Phi_{G}$. So, it will be difficult to demonstrate that $\Phi_{G}=0$ in this case by just employing (\ref{gphase in noncommute}) here. However, it turns out that, this non-vanishing
$\delta^{(2)}(t)$ can be eliminated by using a time-independent unitary
transformation $\boldsymbol{U}\in SU(1,1)\otimes U(1)$. Indeed, by making use of the time-independent unitary transformation
(\ref{unitary}) and the relations in (\ref{equival}), we readily see from (\ref{nch}) that
 \begin{equation}
 \mathcal{H}^{(2)}(t)=\boldsymbol{U}\mathcal{H}^{(1)}(t)\boldsymbol{U}^{\dagger}
 \label{funda}
 \end{equation}
 where the respective parameters are given in (\ref{alph}) and $\beta=\beta_{1} $ in (\ref{beta1}). This indicates that $U(1)$ part of the total dynamical symmetry group $SU(1,1)\otimes U(1)$ (\ref{algcom}) (See Appendix-B) is also fixed by this fine tuned value of $\beta=\beta_{1}$ in (\ref{beta1}); it is not arbitrary. This, in turn, fixes all other parameters of $SU(1,1)$ in (\ref{alph}). And this feature makes it difficult to demonstrate the vanishing of $ \Phi_{G}$, just by employing (\ref{gphase in noncommute}), as mentioned above. In any case, we see that the dilatation term in (\ref{dc}) can be eliminated for all
time $t$ and thereby ensuring $\Phi_{G}=0$. Thus, for this particular value of $\xi=\xi_{c}$, the phase turns out to be integrable. Perhaps, a more transparent way to understand it, would be to consider the following identity

\begin{equation}
 \alpha^{(2)}(t)\beta^{(2)}(t)-\left(\delta^{(2)}(t)\right)^{2}=\alpha^{(1)}(t)\beta^{(1)}(t)~~~~ \forall t
 \label{SUP}
 \end{equation}
 
which follows trivially from (\ref{eq:alpha exp2},\ref{eq:alpha exp3}) and can
			be regarded as a corollary of (\ref{funda}). This demonstrates the invariance under the $SU(1,1)$ or rather  $SO(2,1)=SU(1,1)/\mathbb{Z}_2$ subgroup of $ SU(1,1)\otimes U(1)$,
			 of the corresponding frequency (\ref{su})(See Appendix-B):
			$\omega(t;\xi_{c})=2\sqrt{\alpha^{(1)}(t)\beta^{(1)}(t)}=2\sqrt{\alpha^{(2)}(t)\beta^{(2)}(t)-(\delta^{(2)}(t))^2}$,
			written in terms of either set of the parameters. This, in turn, implies that the
			parameters are indeed connected by $SO(2,1)$ transformation
			(\ref{sop},\ref{su}), and as has been elaborated in Appendix B, this
			$\delta^{(2)}(t) $ can be regarded as the time component of a space-like
			3-vector. But now it can be eliminated for all time by a  global (time-independent) "Lorentz
			transformation", in (2+1)D. And finally when this $SO(2,1) $ transformation
			matrix is lifted to its covering group $SU(1,1)$ ( See Appendix-B), it gives
			the $ SU(1,1)$ part of the transformation matrix $\textbf{U}\in SU(1,1)\otimes
			U(1)$ (\ref{unitary}) (see Appendix-A).


Since this $\boldsymbol{U}$ is time-independent, no connection term of the
manner ((\ref{connection}) in Appendix-B)  will arise here, and
for any state $\left|\Psi(t)\right\rangle $   whose time evolution is governed
by $ \mathcal{H}^{(2)}(t)$ as
$i\hbar\partial_{t}\left|\Psi(t)\right\rangle=\mathcal{H}^{(2)}(t)\left|\Psi(t)\right\rangle$,
we have a corresponding state $
\left(\boldsymbol{U}^{\dagger}\left|\Psi(t)\right\rangle\right)$, which time-evolves by $\mathcal{H}^{(1)}(t)$, as:
$i\hbar\frac{\partial\left(\boldsymbol{U}^{\dagger}\left|\Psi(t)\right\rangle\right)}{\partial t}=\mathcal{H}^{(1)}(t)\left(\boldsymbol{U}^{\dagger}\left|\Psi(t)\right\rangle\right)$.
And thus we are back with $\mathcal{H}^{(1)}(t)$ again, where the corresponding $\delta^{(1)}(t)=0$ and the Berry connection
vanishes as a result. In other words, for this critical value $\xi_{c}$
of the parameter $\xi$, it is indeed possible to transform away the dilatation
term, by subjecting the Hamiltonian $\mathcal{H}^{(2)}(t)$ to a time-independent
unitary transformation $
\mathcal{H}^{(2)}(t)\rightarrow\boldsymbol{U}^{\dagger}\mathcal{H}^{(2)}(t)\boldsymbol{U}$
(\ref{funda}), resulting in vanishing of the Berry phase.\\

This analysis of course will not hold for any other values of $\xi$
other than $\xi_{c}$ i.e for $\xi\ne\xi_{c}$. Nevertheless, here also, it is
possible to eliminate the dilatation term completely, but only by a unitary
transformation $\mathcal{W}(t) \in SU(1,1)$ (\ref{W}) which is neccessarily
time-dependent. Here, unlike the the above case, it is not essential however, to
have $\mathcal{W}(t)$ to be belonging to the entire product group $
SU(1,1)\otimes U(1)$; retaining $ U(1)$ part becomes optional. Consequently
the disappearance of Berry's phase here is only apparent in nature
\cite{doi:10.1142/S0217751X94001047,doi:10.1142/S0217751X88000114} and it gets
lodged in the dynamical part, albeit retaining its geometrical characteristics
\cite{Giavarini:1988wy}. This has been elaborated in Appendix-B.

\end{itemize}

Now returning back to our main objective, let us try to relate this geometric phase shift obtained in Heisenberg picture,
with the more familiar form of Berry phases acquired by state vectors. For that we revert back to the Schrodinger picture. First let's rewrite (\ref{eq:a+sol}) and (\ref{eq:a-sol}) as,

\begin{equation}
\boldsymbol{a}_{\pm}(T)=\boldsymbol{a}_{\pm}(0)\exp\left(-i\Theta_{\pm,d}-i\Phi_{G}\right)
\label{apmmt}
\end{equation}

where

\begin{equation}
\Theta_{\pm,d}=\int_{0}^{T}\left(\hslash\omega\mp\gamma\hslash\right);\Phi_{G}=\int_{0}^{T}\frac{\beta}{\omega}\frac{d}{d\tau}\left(\frac{\delta}{\beta}\right)d\tau\label{fp}
\end{equation}

are the dynamical and geometric phases respectively.

Let $\mathcal{U}(0,t)$ be the Schrodinger evolution operator of our
concerned system, generated by the Hamiltonian (\ref{comh}). Then, $\boldsymbol{a}_{\pm}(t)=\mathcal{U}^{\dagger}(0,t)\boldsymbol{a}_{S\pm}(t)\mathcal{U}(0,t)$,
where $\boldsymbol{a}_{S\pm}(t)$ are the ladder operators in Schrodinger
picture. Note that the time dependence is not entirely frozen here,
even in this Schrodinger picture; it creeps in through the time dependent
parameters.\\

We can therefore write,

\begin{equation}
\begin{split}
&\frac{\left(\boldsymbol{a}_{+}^{\dagger}(T)\right)^{n_{1}}\left(\boldsymbol{a}_{-}^{\dagger}(T)\right)^{n_{2}}}{\sqrt{n_{1}!}\sqrt{n_{2}!}} \left|0,0;t=0\right\rangle _{S}\\
 &=\mathcal{U}^{\dagger}(0,T)\frac{\left(\boldsymbol{a}_{S+}^{\dagger}(T)\right)^{n_{1}}\left(\boldsymbol{a}_{S-}^{\dagger}(T)\right)^{n_{2}}}{\sqrt{n_{1}!}\sqrt{n_{2}!}}\mathcal{U}(0,T)\left|0,0;t=0\right\rangle _{S}\\
&=\mathcal{U}^{\dagger}(0,T)\frac{\left(\boldsymbol{a}_{S+}^{\dagger}(T)\right)^{n_{1}}\left(\boldsymbol{a}_{S-}^{\dagger}(T)\right)^{n_{2}}}{\sqrt{n_{1}!}\sqrt{n_{2}!}}e^{-i\phi_{0,0}}\left|0,0;t=T\right\rangle _{S}\\
&=\mathcal{U}^{\dagger}(0,T)\left|n_{1},n_{2};t=T\right\rangle _{S}e^{-i\phi_{0,0}}\\
&=e^{i(\phi_{n_{1},n_{2}}-\phi_{0,0})}\left|n_{1},n_{2};t=0\right\rangle _{S}
\end{split}
\label{cb1}
\end{equation}

where, $\phi_{n_{1},n_{2}}$ represents the total adiabatic phase
acquired by $\left|n_{1},n_{2};t=0\right\rangle _{S}$ after evolving
by $\mathcal{H}(t)$ over its complete period $T$. Further using (\ref{apmmt}) we also find, 

\begin{equation}
\begin{split}
&\frac{\left(\boldsymbol{a}_{+}^{\dagger}(T)\right)^{n_{1}}\left(\boldsymbol{a}_{-}^{\dagger}(T)\right)^{n_{2}}}{\sqrt{n_{1}!}\sqrt{n_{2}!}} \left|0,0;t=0\right\rangle _{S}\\
 & =e^{in_{1}\left(\Theta_{+,d}+\Phi_{G}\right)}e^{in_{2}\left(\Theta_{-,d}+\Phi_{G}\right)}
 \frac{\left(\boldsymbol{a}_{+}^{\dagger}(0)\right)^{n_{1}}\left(\boldsymbol{a}_{-}^{\dagger}(0)\right)^{n_{2}}}{\sqrt{n_{1}!}\sqrt{n_{2}!}}\left|0,0;t=0\right\rangle _{S}\\
 & =e^{i n_{1}\left(\Theta_{+,d}+\Phi_{G}\right)}e^{i n_{2}\left(\Theta_{-,d}+\Phi_{G}\right)}\left|n_{1},n_{2};t=0\right\rangle _{S}.
\end{split}
\label{cb2}
\end{equation}

Note that here we have made use of the fact that $\boldsymbol{a}_{\pm}(t=0)=\boldsymbol{a}_{S\pm}(t=0).$
Now comparing the above two equations (\ref{cb1}) and (\ref{cb2}),
we get

\begin{equation}
\phi_{n_{1},n_{2}}=\phi_{0,0}+\left[n_{1}(\Theta_{+,d}+\Phi_{G})+n_{2}(\Theta_{-,d}+\Phi_{G})\right]\label{eq:adiaphase}
\end{equation}

So, the Berry phase acquired by the state $\left|n_{1},n_{2};t=0\right\rangle _{S}$
is given by,

\begin{equation}
\phi_{B}^{(n_{1},n_{2})}=\phi_{B}^{(0,0)}+(n_{1}+n_{2})\Phi_{G}
\end{equation}

This kind of linear nature in the Berry phases of different eigenstates,
is a general result \cite{PhysRevA.55.869} for any Hamiltonian with
equally spaced discrete spectrum. And in our case, the total Hamiltonian
(\ref{dh}) is partitioned into two commuting parts corresponding
to $\boldsymbol{a}_{+}$ and $\boldsymbol{a}_{-}$, where each part
produces its own equally spaced spectrum in their respective sub-Hilbert
spaces $\mathbb{H}_{\pm}$ and whose tensor product forms the total Hilbert
space: $\mathbb{H}=\mathbb{H}_{+}\otimes\mathbb{H}_{-}.$\\

Importantly, it is the difference of the Berry phases of different
eigenstates, which contributes in the expectation value of any operator
at time $t$ in a state obtained from any initial state and evolving
under an adiabatic Hamiltonian, $\left\langle \hat{\boldsymbol{O}}\right\rangle (t)=\left\langle \psi(t)\left|\hat{\boldsymbol{O}}(t)\right|\psi(t)\right\rangle $,
where the ground state contribution $\phi_{B}^{(0,0)}$ cancels out.
And most experiments concerning Berry's phase \cite{berryHeis} employ
this idea only. Our derivation certainly provides complete information which, in principle,
may facilitate the predictions of such cases.


\section{Classical analogue: Hannay angles}

We now take-up the study of classical analogue of this quantal geometric
phase, namely the Hannay angles \cite{Hannay_1985}. To clinch the
correspondence we will exploit the correspondence principle of quantum
mechanics with classical mechanics, using coherent states \cite{Maamache_1990}
and some suitable chosen quantum operators that represents the classical
action and angle variables. For that, it will be convenient to recall the concept of instantaneous Hamiltonians introduced in the previous section to discuss the concept of time-reversal symmetry in our context and to  interpret the time-dependent Hamiltonian $\mathcal{H}(t)$ (\ref{comh}), whose time dependence stems from the presence of a set of time dependent parameters $\alpha(t),\beta(t),\delta(t)$ and $\gamma(t)$, as a collection of infinite number of time independent planar systems labeled by time say $t_{0}$ and the set of parameters take their values to be fixed by their respective values for time $t_{0}$  as $\alpha(t_{0}),\beta(t_{0})
,\gamma(t_{0}),\delta(t_{0})$. Left on their own, these individual systems evolving by Hamiltonians like $\mathcal{H}(t_{0})$, with parametric values frozen at $\alpha(t_{0}),\beta(t_{0})
,\gamma(t_{0}),\delta(t_{0})$, will give rise to periodic motion in their respective phase-spaces at classical level, as we show below.  In fact, each such instantaneous Hamiltonians $\mathcal{H}(t_{0})$ can be brought to the standard form of a pair of de-coupled planar oscillators by suitable unitary (resp. canonical) transformations of the respective quantum (resp. classical) systems. And, this ensures the occurence of periodic motion in the phase-spaces, facilitating the introduction of canonical action and angle variables, for each of these classical systems corresponding to the instantaneous Hamiltonians. As one can easily see that we can accomplish this task by introducing  a set of canonically conjugate position and momentum
operators, the so-called quadrature variables, from the ladder operators $\boldsymbol{a}_{+}$ and $\boldsymbol{a}_{-}$
as,



\begin{equation}
\begin{alignedat}{1}\hat{q}_{\pm}= & \sqrt{\frac{\hslash\alpha}{2\omega_{\pm}}}(\boldsymbol{a}_{\pm}^{\dagger}+\boldsymbol{a}_{\pm})\\
\hat{p}_{\pm}= & i\sqrt{\frac{\hslash\omega_{\pm}}{2\alpha}}(\boldsymbol{a}_{\pm}^{\dagger}-\boldsymbol{a}_{\pm})
\end{alignedat}
\label{quadr}
\end{equation}

Utilizing (\ref{ol},\ref{eq:J2 to J3}), this shows $\hat{q}_\pm=\hat{q}_\pm(q_{1},q_{2},p_{1},p_{2}),\hat{p}_\pm=\hat{p}_\pm(q_{1},q_{2},p_{1},p_{2})$ are linearly dependent on the old coordinates and momenta. At the quantum level, this can be implemented at each such insntants $t_0$, by suitable unitary transformations $\mathcal{V}(t_0)\in SU(1,1)\otimes U(1)$. This is exactly like the case of $\mathcal{W} (t_0)$ as in (\ref{W}) (see Appendix B), which helped us to eliminate just the dilatation term, without touching the Zeeman term in  $\mathcal{H}(t)$ (\ref{comh}), except that we are now eliminating the Zeeman term also. But the explicit construction of such a $\mathcal{V}(t_0)$ is neither easy nor required in our case, as we are dealing with the instantaneous classical systems here. In fact, as one can easily see, at the classical level (where phase-space variables are just c-numbers) an analogues linear canonical transformation with the coefficients determined by the values of $\alpha,\beta,\gamma,\delta$ at $t=t_0$, canonically transforms the instantaneous classical systems from \{$q_1,q_2;p_1,p_2$\} canonical pairs to \{$q_{+},q_{-};p_{+},p_{-}$\} canonical pairs. Correspondingly, we get from the classical Hamiltonian in old phase-space variables, which is the classical counterpart of our quantum Hamiltonian (\ref{comh}) at the instant $t_0$ , the 2D decoupled harmonic oscillator like Hamiltonian, written just in terms of new phase-space variables :

\begin{equation}
\mathcal{H}_{Cla} (t_0)=\frac{\omega_{+}^{2}}{2\alpha}q_{+}^{2}+\frac{\alpha}{2}p_{+}^2+\frac{\omega_{-}^{2}}{2\alpha}q_{-}^{2}+\frac{\alpha}{2}p_{-}^2.
\label{cla}
\end{equation}

This is the classical counterpart of unitarily transformed Hamiltonian
 (\ref{comh}): $\mathcal{V}(t_0)\mathcal{H}(t_0)\mathcal{V}^{\dagger}(t_0)$. Now if we were to consider time-evolution in the full time-dependent system governed by $\mathcal{H}(t)$, instead of just the time-independent Hamiltonians $\mathcal{H}(t_0)$ with parameters frozen at fixed values, we would have required to augment  the unitary transformed Hamiltonian  by a suitable connection term like $i\hbar\mathcal{V}(t)\partial_{t}\mathcal{V}^{\dagger}(t) $, so that, the total Hamiltonian $\mathcal{H}_{total}(t)$, as in (\ref{connection}), can govern the time evolution of the transformed states ($\mathcal{V}(t)\left|\Psi(t)\right\rangle $). And as shown in the case of $\mathcal{W} (t)$ (\ref{W}) in Appendix-B, here too we can show that the geometrical phase will occur in the dynamical phase obtained through $\mathcal{H}_{total}(t)$, but will retain its geometrical feature. Similarly the classical counterpart of this $\mathcal{H}_{total}(t)$ can be obtained by simply adding a term of the form $\frac{\partial F}{\partial t}$, where $F$ is a suitable generating function \cite{PhysRevA.43.5717}, to (\ref{cla}).
 In this section, we are, of course, not bothered about this extra time-derivative term in $\mathcal{H}_{total}(t)$, because the parameters $\alpha$, $\beta$, $\gamma$, $\delta$ are held frozen to their respective values corresponding to the instant $t=t_0$. Besides, each such instantaneous Hamiltonians $\mathcal{H}_{Cla}(t_0)$ (\ref{cla}) in the classical case, give rise to periodic motion in phase-space, if considered as a separate system on its own, and hence allows us to introduce corresponding action and angle variables, as mentioned above.\\\\
Now in this classical case \cite{Maamache_1990}, let $\{C(I,\boldsymbol{R})\}$
 denotes a continuous family of periodic trajectories $C(I,\boldsymbol{R})$
 in the phase space associated with the classical Hamiltonians $\mathcal{H}_{Cla}(\boldsymbol{R})$
 and let $\omega(I,\boldsymbol{R})$ be the angular velocity on $C(I,\boldsymbol{R})$,
 where each curve is equipped with a definite origin for angle
 variable $v$ conjugated to the action variables $I$. Now, during the adiabatic evolution, a point in phase
 space follows a trajectory of constant action, and only the  angular
 coordinate $v(t)$ will evolve in time and its value at time $t$  is given by:

\begin{equation}
v(t)=v(0)+\int_{0}^{t}\omega(I,\boldsymbol{R}(s))\mathrm{d}s+\Delta v_{I}^{\mathrm{H}}(t).
\end{equation}

This involves as an integration along the curve $C(I,\boldsymbol{R}(t))$,
and also contains, apart from the usual dynamical contribution, a
geometrical one also, the so-called Hannay's angle $\Delta v_{I}^{\mathrm{H}}(t)$.
Note that, since our classical Hamiltonian $\mathcal{H}_{Cla}(\boldsymbol{R})$
has two degrees of freedom, so we expect two sets of action-angle
coordinates $\{\{I_{i},v_{i}\}:i=1,2\}$.\\

Now let's consider the coherent states\cite{PhysRev.131.2766,Gazeau2000,Novaes_2002}
of our two-dimensional harmonic oscillator (\ref{dh}), which are
supposed to be the best approximations to a classical state. The coherent
states analogous to \cite{Li_2018} in this case are the tensor product
of two independent Glauber-Klauder-Sudarshan (GKS)-coherent states
\cite{Novaes_2002}, which are simultaneous (normalised) eigen states
of the two mutually commuting annihilation operators $\boldsymbol{a}_{+}$
and $\boldsymbol{a}_{-}$:

\begin{equation}
\begin{alignedat}{1}\left|z_{1},z_{2};\boldsymbol{R}\right\rangle = & \left|z_{1}(\boldsymbol{R})\right\rangle \otimes\left|z_{2}(\boldsymbol{R})\right\rangle \\
\boldsymbol{a}_{+}\left|z_{1}(\boldsymbol{R})\right\rangle = & z_{1}\left|z_{1}(\boldsymbol{R})\right\rangle \\
\boldsymbol{a}_{-}\left|z_{2}(\boldsymbol{R})\right\rangle = & z_{2}\left|z_{2}(\boldsymbol{R})\right\rangle \\
\left|z_{1},z_{2};\boldsymbol{R}\right\rangle = & \mathrm{e}^{-\left(\left|z_{1}\right|^{2}+\left|z_{2}\right|^{2}\right)/2}\sum_{n_{1}=0}^{\infty}\sum_{n_{2}=0}^{\infty}\frac{z_{1}^{n_{1}}}{\sqrt{n_{1}!}}\frac{z_{2}^{n_{2}}}{\sqrt{n_{2}!}}\left|n_{1},n_{2};\boldsymbol{R}\right\rangle 
\end{alignedat}
\end{equation}

Further it has been shown in \cite{Maamache_1990}, that a suitable
quantum operator for classical action $I_{i}$  is $\hat{I}_{i}(\boldsymbol{R})=\hslash\hat{N}_{i}(\boldsymbol{R})$,
where $\hat{N}_{i}(\boldsymbol{R})$ is the $i$-th number operator.
with $i\in\{+,-\}$ in our case. Now, let $\hat{U}_{i}(\boldsymbol{R})$'s
be the unitary operators defined through their action, $\hat{U}_{1}(\boldsymbol{R})\left|n_{1},n_{2};\boldsymbol{R}\right\rangle =\left|n_{1}-1,n_{2};\boldsymbol{R}\right\rangle $;
$\hat{U}_{1}(\boldsymbol{R})\left|0,n_{2};\boldsymbol{R}\right\rangle =0$
and similarly for $\hat{U}_{2}(\boldsymbol{R})$. They essentially
correspond to the well known polar decompositions of the operators like 
$\boldsymbol{a}$ into the so-called number $\hat{N}$ and phase
operators $\hat{\theta}$: $\boldsymbol{a}_{\pm}=\sqrt{\hat{N}_{\pm}}e^{i\hat{\theta}_{\pm}}$,
with $\hat{U}_{i}(\boldsymbol{R})$ can be thought of the operator
corresponding to $e^{-i\hat{\theta_{i}}}.$ It is also shown that the expectation
values of these operators in the state $\left|z_{1},z_{2};\boldsymbol{R}\right\rangle $
are given by, $I_{i}=\left\langle \hat{I}_{i}(\boldsymbol{R})\right\rangle =\left|z_{i}\right|^{2}\hslash$
and $\left\langle \hat{U}_{i}(\boldsymbol{R})\right\rangle =e^{i\times arg(z_{i})}$,
so that in the classical limit, we can identify $z_{j}=\sqrt{\frac{I_{j}}{\hslash}}e^{-iv_{j}}$.\\

And this natural relationship between the ladder operators of the quantum system and the corresponding action and angle-like operators was the main driving motivation behind our unconventional approach to determine the Berry phases by solving the evolution-equations of $\boldsymbol{a}_{\pm}$ in Heisenberg picture, which also provides a natural framework for semiclassical correspondence. In fact, the geometrical part of the phases acquired by $\boldsymbol{a}_{\pm}$ over a complete period of adiabatic-cycle, as found in (\ref{eq:a+sol},\ref{eq:a-sol}), is precisely the Hannay's angle of the corresponding classical adiabatic evolution, as we identify below. So we see that we could determine at one go, the Berry phases as well as the Hannay's angle from (\ref{eq:a+sol},\ref{eq:a-sol}). Besides, had we taken the more conventional route we would have required to determine the exact energy eigen-functions of the quantum system in order to find the intended geometric phases, which is not a very easy job to do for a generalized two-dimesional Harmonic oscillator such as (\ref{comh}). Hence the overall approach we took, though was not the generic one, suited our desired goals more approprately.\\

Now returning back to the original point, we now consider the following wave packet as the initial state, which best approximates the initial conditions of the corresponding classical adiabatic evolution,

\begin{equation}
\left|z_{1},z_{2};\boldsymbol{R}(0)\right\rangle =\mathrm{e}^{-\left(\left|z_{1}\right|^{2}+\left|z_{2}\right|^{2}\right)/2}\sum_{n_{1}=0}^{\infty}\sum_{n_{2}=0}^{\infty}\frac{z_{1}^{n_{1}}}{\sqrt{n_{1}!}}\frac{z_{2}^{n_{2}}}{\sqrt{n_{2}!}}\left|n_{1},n_{2};\boldsymbol{R}(0)\right\rangle 
\end{equation}
and evolve it adiabatically over a complete cycle, to get, using (\ref{eq:adiaphase})

\begin{equation}
\begin{aligned}\mathcal{U}(0,T)\left|z_{1},z_{2};\boldsymbol{R}(0)\right\rangle = & \mathrm{e}^{-\left(\left|z_{1}\right|^{2}+\left|z_{2}\right|^{2}\right)/2}\sum_{n_{1}=0}^{\infty}\sum_{n_{2}=0}^{\infty}\frac{z_{1}^{n_{1}}}{\sqrt{n_{1}!}}\frac{z_{2}^{n_{2}}}{\sqrt{n_{2}!}}e^{-i\phi_{0,0}}\left|n_{1},n_{2};\boldsymbol{R}(T)\right\rangle \times\\
 & e^{-in_{1}(\Theta_{+,d}+\Phi_{G})}e^{-in_{2}(\Theta_{-,d}+\Phi_{G})}\\
= & \mathrm{e}^{-\left(\left|z_{1}\right|^{2}+\left|z_{2}\right|^{2}\right)/2}\times e^{-i\phi_{0,0}}\sum_{n_{1}=0}^{\infty}\sum_{n_{2}=0}^{\infty}\frac{\left(z_{1}e^{-i(\Theta_{+,d}+\Phi_{G})}\right)^{n_{1}}}{\sqrt{n_{1}!}}\times\\
 & \frac{\left(z_{2}e^{-i(\Theta_{-,d}+\Phi_{G})}\right)^{n_{1}}}{\sqrt{n_{2}!}}\left|n_{1},n_{2};\boldsymbol{R}(T)\right\rangle \\
= & e^{-i\phi_{0,0}}\left|z_{1}e^{-i(\Theta_{+,d}+\Phi_{G})},z_{2}e^{-i(\Theta_{-,d}+\Phi_{G})};\boldsymbol{R}(T)\right\rangle 
\end{aligned}
\label{eq:coherent evolution}
\end{equation}

Therefore, up to a global phase factor, the coherent
state associated with the initial Hamiltonian $\mathcal{H}(0)$, evolves to another coherent state $\left|z_{1}(T),z_{2}(T);\boldsymbol{R}(T)\right\rangle $, now
associated with the Hamiltonian $\mathcal{H}(T)$ at time T, where
$z_{1}(T)=z_{1}e^{-i(\Theta_{+,d}+\Phi_{G})}$ and $z_{2}(T)=z_{2}e^{-i(\Theta_{-,d}+\Phi_{G})}$. Thus, from the evolution of $z_{1}$ and $z_{2}$ in (\ref{eq:coherent evolution}),
through a complete period of the adiabatic Hamiltonian, we can identify
$\Theta_{\pm,d}=\int_{0}^{T}\omega_{\pm}(t')dt'$, where $\omega_{i}=\frac{1}{\hslash}\frac{\partial E_{n_{1},n_{2}}}{\partial n_{i}}$(Like
$\frac{\partial\mathcal{H}_{Cla}}{\partial I_{i}}$), with the dynamical
phases and $\Phi_{G}$ (from (\ref{eq:ladderphase}))
with the angular shift which was obtained classically by Hannay.

The expectation values of the new set of position-momentum i.e. the
quadrature operators (\ref{quadr}), which are now time-dependent, are found to be given by\cite{PhysRev.131.2766},

\begin{equation}
\begin{alignedat}{1}\left\langle \hat{q}_{\pm}\right\rangle = & \sqrt{2\hslash/\omega_{\pm}}Re(z_{i})\\
\left\langle \hat{p}_{\pm}\right\rangle = & \sqrt{2\hslash\omega_{\pm}}Im(z_{i})\\
 & \text{(i=1,2 respectively)}
\end{alignedat}
\label{scrodinger}
\end{equation}

On adopting the parametrization of $z_{1}$ and $z_{2}$, introduced
above, the expectation values of these phase space operators in the
transported state are obtained as,

\begin{equation}
\begin{alignedat}{1}\left\langle \hat{q}_{\pm}\right\rangle _{T}= & \sqrt{2I_{i}/\omega_{\pm}}Cos(v_{i}(0)+\Theta_{\pm,d}+\Phi_{G})\\
\left\langle \hat{p}_{\pm}\right\rangle _{T}= & -\sqrt{2I_{i}\omega_{\pm}}Sin(v_{i}(0)+\Theta_{\pm,d}+\Phi_{G}),
\label{qhan}
\end{alignedat}
\end{equation}

showing that the corresponding classically canonical conjugate phase space variables i.e the classical counterpart of quadrature operators 
$\hat{q}_{\pm},\hat{p}_{\pm}$ (\ref{quadr}) undergo oscillatory motion. Thus the geometric
phase $\Phi_{G}$,  entering into the non-stationary coherent-state through all of  its stationary components i.e. the energy eigenstates, generates in the classical limit ($\hbar \rightarrow 0$ ,$\lvert z_{i} \lvert \rightarrow \infty$ , $ \sqrt{I_{i}} =\sqrt{\hbar} \lvert z_{i} \lvert \rightarrow$ finite), the angle variable conjugate to the  action $I_{i} $ and are given by the phase of $ z_{i} $. Therefore the additional phase of $ z_{i} $ i.e one over and above the dynamical phase, can be identified with Hannay angle,  which can clearly be  understood from classical arguments.\\

Before we conclude this section, we would like to draw the attention of the reader to an earlier work \cite{berryHeis}, where  also Berry phase was computed using the Heisenberg picture. However, in contrast to our approach, they had incorporated
	adiabatic approximation right at the level of time-dependent Schrodinger equation, thereby identifying an effective adiabatic Hamiltonian $\mathcal {H}_{ad}(t)$
	as described in \cite{book:1722}\cite{PhysRevA.44.5383}, which then governs the time evolution of adiabatic state vectors in Schrodinger picture. That also helped them to obtain an effective Heisenberg equation of motion for any general phase-space operator, under adiabatic approximation. On the other hand, in our approach, we worked with the original Hamiltonian itself and only  implemented adiabaticity while solving the respective Heisenberg equations of motion explicitely.\\\\
Finally, to clinch the correspondence with Hannay's angle, the authors of \cite{berryHeis} directly took expectation of the time-evolved quantum operator calculated in the way mentioned above, with the initial coherent state of the system best suited for the initial conditions of the dynamics. And finally, extracted the Hannay's angle from the oscillatory sinusoidal terms, which resembles the ones we had in (\ref{scrodinger}), occuring in the time-evolved expectation-value obtained in Heisenberg-picture.
	This is again in contrast to our case, as we determined the Hannay's angle of our system using Schrodinger-picture. That is, we calculated the time-evolved expectation value, using initial ($t=0$) quantum operator and time-evolved coherent state, after a complete periodic cycle of the Hamiltonian. On the way of course, we had to make use of our preceding findings, namely the Berry phases calculated in Heisenberg picture. Further, while determining the Hannay's angle from time-evolved expectation, we made use of two special quantum operators \cite{Maamache_1990}, which are the authentic quantum versions of the classical action and angle variables, and thereby begetting
	the genuine classical correspondence with Hannay's angle.

 \section{ Conclusions}
 We have considered the system of harmonic oscillator in  Moyal plane, but with the additional feature that there is noncommutativity among momentum components also like the spatial ones and other parameters are varying slowly with time. Although periodicity, rather that adiabaticity, is more relevant in the computation of geometrical phase, as shown by Aharanov and Anandan \cite{PhysRevLett.58.1593}, we nevertheless find the original adiabatic approach due to Berry convenient to execute. For that we introduce a novel form of Bopp shift, which is more general in nature and does not involve any effective Plank constant $ \hbar_{eff} $, as has been done in the literature . Through this Bopp shift, we can generate a certain dilatation term involving commutative phase space variable ($[q_{i},q_{j}]=0=[p_{i},p_{j}];[q_{i},p_{j}]=i\hslash\delta_{ij} $), which plays an indispensable role in generating this geometrical i.e Berry phase.We have also provided a novel and  unconventional approach to compute this geometrical phase shift initially in Heisenberg picture and then related it with the conventional Berry phase in the Schrodinger picture. Finally, the classical analogue of Hannay angle was also computed using Glauber-Sudarshan coherent states. We finally observe that the emergent Berry phase( geometrical phase shift) depends on  both types of noncommutative parameters ($\theta$ and $ \eta$) and it  will  vanish in the situation if either one of these parameters were to vanish.  Thus we can conclude that, the noncommutative phase space structure  induces a suitable geometry on the  circuit $ \Gamma $ in parameter space of the  2D time dependent  harmonic oscilator system , which manifests in the appearance of the associated geometrical phase shift, when a circuital adiabatic excursion in the parameter space is considered. \\

We would also like to mention that, the effective commutative
			Hamiltonian $\mathcal{H}(t)$ (\ref{comh}), obtained by making use of our Bopp
			shift (\ref{bpshift}) takes its value in $su(1,1)\oplus u(1)$  Lie Algebra (\ref{gro})
			which contain terms, responsible for explicit breaking of time-reversal symmetry
			of the family of instantaneous Hamiltonians $\mathcal{H}(t)$'s, which is a
			necessary condition to get the Berry's phase \cite{PhysRevLett.67.251,j.Ihm}.
			The instantaneous eigenstates of the Hamiltonians $\mathcal{H}(t)$ can therefore
			be taken to belong to the representation space of the group $SU(1,1)\otimes
			U(1)$ and the occurence of geometrical phase becomes inevitable in this case,
			as has been shown in \cite{Dutta_Roy_1993,RevModPhys.62.867}. Our result
			therefore corroborates this general observation. And as long as, the noncommutative parameters $\theta$ and $\eta$ can be regarded 				as fundamental parameters in some appropriate energy scale, the resulting Berry phase can also be regarded as fundamental.\\
			

		Finally regarding physical models exhibiting Berry phase with phase space
			noncommutative structures, one can perhaps envisage designing a planar system of charged anyons, having fractional spin (related to $\theta$) and trapped in a harmonic potential well, with a normal magnetic field $B$ (related to $\eta$) \cite{NAIR2001267,BELLUCCI2002295,PhysRevD.65.107701} and where the mass and frequency parameters are both taken to be slowly varying as periodic functions of time.  We feel that with this, one can have some experimental demonstrations of this phenomena and which should have some bearings in condensed matter systems \cite{book:174801,Jellal_2001}.\\

 
Last but not the least, we briefly mention some interesting directions in which our present work can be extended. The first is to construct  coherent state Euclidean path integral \cite{Kashiwa:1992np,PhysRevLett.102.241602} formulation invoking adiabatic  iterative prescription \cite{doi:10.1098/rspa.1987.0131,PhysRevA.40.526}  for
 calculating non-adiabatic corrections on Berry phase in non-commutative phase space. Apart from 2D oscillator, one also can think of other exactly integrable system where the partition function in the coherent state path integral method can be computed so that one can  eventually obtain the Gibb's entropy of the system and also try to make a possible connection with von Neumann entropy\cite{PhysRevLett.71.666,Kabat:1994vj} of the system in presence of the  modified geometric phase.\\
 
  The analysis presented here can perhaps also  be extended to compute quantum information metric  and Berry curvature \cite{Alvarez-Jimenez:2017gus} from the effective action \cite{PhysRevLett.102.241602} corresponding to the above mentioned path integral, so that one can try to connect some features in noncommutative quantum mechanics with quantum information science. Of course, any endeavor to detect the effect of noncommutativity is a challenging enterprise. In light of the
 present paper, there could appear many surprises in this area. We hope to return to some of these
 issues in our future work(s).

\section*{Acknowledgements}

One of the authors (P.N.)  thanks  Professor M. V. Berry for  enlightening him about quite a few subtle points in connection with this paper. Also he has benefited from conversation with Sayan K.Pal. S.B., would like to thank the authorities of S.N Bose Centre for providing  a very pleasant atmosphere, academic or otherwise, and their kind hospitality during the course of his stay in the S.N. Bose National Centre for Basic Sciences when this work was initiated and also would like to thank KVPY-India for providing financial support in the form of fellowship during the course of this work. Finally, it is a pleasure to thank both the referees for their constructive comments on the earlier version of the draft and we believe that it has helped us to enhance the quality of the paper quite substantially. 



\section *{ Appendix A. On unitary equivalence of different realizations of NC phase
	space algebra:}

As we have already mentioned, there is yet another realisation of the whole phase-space non-commutative algebra
given in \cite{PhysRevLett.93.043002, Bertolami:2005ud}, unlike the one used by us (\ref{bpshift}). 
And Berry phase too was studied using that realisation in non-commutative phase-space, albeit
in a completely different system involving gravitational
potential well \cite{Bastos:2006kj}, but was found to vanish, where an equivalent scaled version of the realisation (\ref{bps1}) \cite{Bertolami:2005ud} given below, was employed.  
What we would like to show here is that, our realisation (\ref{bps3}) is a more general one in the sense that the one parameter ($\xi$) familty of non-commutative algebra (\ref{ncms}) given below and
the realisation (\ref{bps1}) occuring in \cite{PhysRevLett.93.043002, Bertolami:2005ud} is unitarily equivalent to our realisation
(\ref{bps3}), only for a particular value of $\xi=\xi_c$ (\ref{bd}) and hence will produce same physical results for this value only.
However,
our realisation (\ref{bps3}) is the only one which persists to be 
valid for other values of $\xi$, i.e. for $\xi\neq\xi_c$ also.

To demonstrate this above mentioned equivalence between the realisations (\ref{bps1},\ref{bps3}) holding only for $\xi=\xi_{c}$ (\ref{bd}), let's consider
the following structure of noncommutativity among the phase space
variables: 
\begin{equation}
\begin{array}{c}
[\hat{x}_{i},\hat{x}_{j}]=i\xi\theta\epsilon_{ij};[\hat{p}_{i},\hat{p}_{j}]=i\xi\eta\epsilon_{ij};[\hat{x}_{i},\hat{p}_{j}]=i\hslash\delta_{ij}\end{array};\theta\eta<0;\label{ncms}
\end{equation}
where $\theta$ and $\eta$ are constant parameters; $\epsilon_{ij}$
is an anti-symmetric constant tensor and $\xi$ being a scaling parameter. We then introduce the commuting
coordinates $q_{i}$ and momenta $p_{i}$ respectively satisfying
the usual Heisenberg algebra: $[q_{i},q_{j}]=0=[p_{i},p_{j}];[q_{i},p_{j}]$ $=i\hslash\delta_{ij}$.
Note that in our notation, these $q_{i}$'s and $p_{i}$'s carry no
over-head hats, in contrast to their noncommutative counterparts ($\hat{x_{i}}$'s
and $\hat{p_{i}}$ 's).\\

\begin{doublespace}
In \cite{PhysRevLett.93.043002}, the realisation in terms of the above
$q_{i}$'s and $p_{i}$'s are given by the following linear transformation:

\begin{equation}
\begin{alignedat}{1}\hat{x}_{i}^{(1)} & =\sqrt{\xi}(q_{i}-\frac{\theta}{2\hslash}\epsilon_{ij}p_{j})\\
\hat{p}_{i}^{(1)} & =\sqrt{\xi}(p_{i}+\frac{\eta}{2\hslash}\epsilon_{ij}q_{j}),
\end{alignedat}
\label{bps1}
\end{equation}

\end{doublespace}
which holds only if 
		\begin{equation}
		\xi=\xi_{c}:=(1+\frac{\theta\eta}{4\hbar^{2}})^{-1};4\hbar^{2}+\theta\eta>0
		\label{bd}
		\end{equation}

		But this realization of the algebra (\ref{ncms}) is not unique
		\cite{NAIR2001267}. 
		Indeed, we provide below another possible realization of (\ref{ncms}) in terms
		of another linear transformation, as
		\begin{equation}
		\begin{alignedat}{1}\hat{x}_{i}^{(2)} &
		=q_{i}-\frac{\xi\theta}{2\hslash}\epsilon_{ij}p_{j}+\frac{\xi\sqrt{-\theta\eta}}{2\hslash}\epsilon_{ij}q_{j}\\
		\hat{p}_{i}^{(2)} &
		=p_{i}+\frac{\xi\eta}{2\hslash}\epsilon_{ij}q_{j}+\frac{\xi\sqrt{-\theta\eta}}{2\hslash}\epsilon_{ij}p_{j},
		\end{alignedat}
		\label{bps3}
		\end{equation}
		The merit of this realization is that it is valid for any value of
		$\xi$ and need not be fixed to the value given in (\ref{bd}). This is unlike
		the one in (\ref{bps1}). Clearly, neither of the transformations (\ref{bps1}) or
		(\ref{bps3}) represent canonical transformations, as they change the basic
		commutator algebra. It is, however, quite obvious that for the value
		of $\xi$ parameter, fine tuned to value in (\ref{bd}), the realizations
		should be unitarily equivalent. We now construct this unitary transformation
		explicitly, which map the  realization (\ref{bps1}) to the  other one 
		(\ref{bps3}).
		To that end, let us make the following ansatz of the unitary operator:

\begin{equation}
\textbf{U}=\exp[-i\frac{\sigma\textbf{D}}{\hbar}]~\exp[-i\frac{\beta \textbf{L}}{\hbar}]~\exp[-i\alpha_{2}\vec{p}^{2}]~\exp[-i\alpha_{1}\vec{q} ^{2}],
\label{unitary}
\end{equation}
\\
where $\textbf{D}=\vec{q}.\vec{p}+\vec{p}.\vec{q}$ and $\textbf{L}=\vec{q}\wedge\vec{p}$ 
are respectively the dilatation and angular momentum operators\footnote{While former represents scalar operator, 
latter represents a pseudo scalar operator in a commutative plane and generates appropriate transformations. It
is also quite well-known that the  three scalar generators ($\textbf{D},\vec{p}^{2},\vec{q}^{2}$)
form a closed $SO(1,2)$ algebra \cite{doi:10.1142/S0217751X88000114},
while $\textbf{L}$ commutes with all of them:$[\textbf{L},\vec{q}^{2}]=[\textbf{L},\vec{p}^{2}]=[\textbf{L},\textbf{D}]=0$. }, and relates these two representations as,

\begin{equation}
\hat{x}_{i}^{(2)}=\textbf{U}\hat{x}_{i}^{(1)}\textbf{U}^{\dagger},\hat{p}_{i}^{(2)}=\textbf{U}\hat{p}_{i}^{(1)}\textbf{U}^{\dagger}
\label{equival}
\end{equation}\\

Note that we have taken the parameters $\sigma$ and $\beta$ to be dimension-less, in contrast to the parameters $\alpha_{1}$ and $\alpha_{2}$ which are dimensionful. Now making use of Hadamard identity we can easily show that,

\begin{equation}
\begin{alignedat}{1}\hat{x}_{i}^{(2)} & =Aq_{i}-B\epsilon_{ij}p_{j}+C\epsilon_{ij}q_{j}+Dp_{i}\\
\hat{p}_{i}^{(2)} & =Ep_{i}+F\epsilon_{ij}q_{j}+G\epsilon_{ij}p_{j}-Hq_{i},
\end{alignedat}
\label{bps2l}
\end{equation}

where 

\begin{equation}
\begin{alignedat}{1}A= & \lambda\sqrt{\xi}[cos(\beta)+\alpha_{1}\theta sin(\beta)],\ \ \ \ \ B=\frac{\sqrt{\xi}}{\lambda}[(\frac{\theta}{2\hbar}-2\alpha_{1}\alpha_{2}\theta\hbar)cos(\beta)+2\alpha_{2}\hbar sin(\beta)]\\
C= & \sqrt{\xi}\lambda[sin(\beta)-\alpha_{1}\theta cos(\beta)],\ \ \ \ \ D=\frac{\sqrt{\xi}}{\lambda}[(\frac{\theta}{2\hbar}-2\alpha_{1}\alpha_{2}\hbar\theta)sin(\beta)-2\alpha_{2}\hbar cos(\beta)]\\
E= & \frac{\sqrt{\xi}}{\lambda}[(1-4\alpha_{1}\alpha_{2}\hbar^{2})cos(\beta)+\eta\alpha_{2}sin(\beta)],\ \ \ \ \ F=\lambda\sqrt{\xi}[\frac{\eta}{2\hbar}cos(\beta)+2\alpha_{1}\hbar sin(\beta)]\\
G= & \frac{\sqrt{\xi}}{\lambda}[(1-4\alpha_{1}\alpha_{2}\hbar^{2})sin(\beta)-\eta\alpha_{2} cos(\beta)],\ \ \ \ \ H=\lambda\sqrt{\xi}[\frac{\eta}{2\hbar}sin(\beta)-2\alpha_{1}\hbar  cos(\beta)].\\
\lambda= & \exp(-2\sigma).
\end{alignedat}\label{abcefg}
\end{equation}

On the other hand , all these eight coefficients $A$ - $H$ in (\ref{abcefg}) can be determined easily by comparing (\ref{bps2l}) with (\ref{bps3}) and is provided below in two segregated clusters:
\ 
\\

\begin{equation}
H=0,\ \ \ \ \ D=0,\ \ \ \ \ A=1,\ \ \ \ \ C=\xi\frac{\sqrt{-\theta\eta}}{2\hbar}\ \ \ \ \ 
\label{ebfgg}
\end{equation}
and,\\
\begin{equation}
B=\xi\frac{\theta}{2\hbar},\ \ \ \ \ E=1,\ \ \ \ \ F=\frac{\xi\eta}{2\hbar},\ \ \ \ \ G=\xi\frac{\sqrt{-\theta\eta}}{2\hbar}\ \ \ \ \ 
\label{evigpg}
\end{equation}

The reason for this segregation is that a simple inspection suggests that we can make use of first three equations in (\ref{ebfgg}) to solve for $\alpha_{1}$, $\alpha_{2}$, and $\lambda$ in terms of the single parameter $\beta$ as,

\begin{eqnarray}
\begin{array}{rcl}
 &  & \alpha_{1}=\frac{\eta}{4\hbar^{2}}tan(\beta)\\
\\
 &  & \alpha_{2}=\frac{\theta}{4\hbar^{2}}[\frac{tan(\beta)}{1+\frac{\theta\eta}{4\hbar^{2}}tan^{2}(\beta)}]\\
\\
 &  & \lambda=[\sqrt{\xi}(cos(\beta)+\alpha_{1}\theta sin(\beta))]^{-1}
\end{array}\label{alph}
\end{eqnarray}

and then this $\beta $ , can be determined by first making use of the fourth equation in (\ref{ebfgg}) to get the following quadratic equation:
 
\begin{equation}
\frac{\xi}{4\hbar^{2}}(-\theta\eta)^{\frac{3}{2}}tan^{2}(\beta)-(\frac{\theta\eta}{2\hbar}-2\hbar)tan(\beta)-\xi\sqrt{-\theta\eta}=0,
\end{equation}

yielding the following two roots for $\beta$:

\begin{equation}
\beta_{1}= tan^{-1}( \sqrt{\frac{-\theta\eta}{4\hbar^{2}}})  \ \ \ \ \ \beta_{2}=-tan^{-1}((\frac{-\theta\eta}{4\hbar^{2}})^{\frac{3}{2}})
\label{beta1}
\end{equation}.

It is now a matter of a lengthy but straightforward computation to verify that only when $\beta_{1}$ from (\ref{beta1}) along with $\alpha_{1}$, $\alpha_{2}$, and $\lambda$ from (\ref{alph}) are substituted to the set of expressions of $B$, $E$,$F$ and $G$ in (\ref{abcefg}) they readily yield the corresponding expression given in (\ref{evigpg}). This therefore provides an explicit demonstration of the unitary equivalence of two realizations (\ref{bps1},\ref{bps3}) for specific values of $\xi=\xi_c$ in (\ref{bd}).  For other values of $\xi$
the realisation (\ref{bps1}) will not hold, in contrast to the realisation
(\ref{bps3}) which persists to hold. In this sense the realisation
(\ref{bps3}) is more general and in this paper, we are basically working
with the algebra (\ref{non-com}) and its realisation (\ref{bpshift}), which are nothing
but the equations (\ref{bps3},\ref{ncms}) themselves with $\xi=1$.\\

\section*{ Appendix B. On the dynamical symmetry group $SU(1,1)\otimes U(1)$ of the Hamiltonian (\ref{comh}) and apparent removability of the Berry phase:}

Following Wei-Norman method \cite{doi:10.1063/1.1703993} we can readily identify
		the Lie algebraic structure \cite{PhysRevA.37.1934} of the Hamiltonian operator (\ref{comh}). To that
		end let us introduce the generators: 
		\begin{equation}
		\boldsymbol{K_{+}}=\frac{iq^{2}_{i}}{2};\boldsymbol{K_{-}}=\frac{ip^{2}_{i}}{2};
		\boldsymbol{K_{0}}=\frac{i(p_{i}q_{i}+q_{i}p_{i})}{4};~~~
		\boldsymbol{L}=\epsilon_{ij}q_{i}p_{j}
		\label{alg}
		\end{equation}
		
		It can now be  shown quite easily that the   $\boldsymbol{K_{\pm}}$, and
		$\boldsymbol{K_{0}}$ and $\boldsymbol{L}$, satisfy the $su(1,1)\oplus u(1)$ Lie
		algebra \cite{book:16453}:
		\begin{equation}
		[\boldsymbol{K_{0}},\boldsymbol{K_{\pm}}]=\pm \hbar \boldsymbol{K_{\pm}};~~
		[\boldsymbol{K_{+}},\boldsymbol{K_{-}}]=-2\hbar\boldsymbol{K_{0}};~~[\boldsymbol{K_{\pm}},\boldsymbol{L}]=[\boldsymbol{K_{0}},\boldsymbol{L}]=0
		\label{algcom}
		\end{equation}
		
		where $\boldsymbol{L}$ is the $u(1)$ generator commuting with all $su(1,1)$
		generators. Upon exponentiating in a suitable manner, like in (\ref{unitary}),
	they will generate all the elements of the group $SU(1,1)\otimes U(1)$.\\
	
	To be more transparent, let us introduce the dimensionless  generators
	$\boldsymbol{T}_{1} , \boldsymbol{T}_{2}$ $ \boldsymbol{T}_{0}$ defined through
	$\boldsymbol{K}_{\pm}$  and $\boldsymbol{K}_{0}$ as,  
	
	\begin{equation}
	\boldsymbol{K}_{+}=\theta (\boldsymbol{T}_{1}+i \boldsymbol{T}_{2});
	\boldsymbol{K}_{-}=-\eta (\boldsymbol{T}_{1}- i\boldsymbol
	{T}_{2});\boldsymbol{T}_{0}=\frac{\boldsymbol{K}_{0}}{\sqrt{-\theta\eta}}; \theta\eta<0
	\label{dimalg}
	\end{equation}
	
	where $ \theta$ and $\eta$ are the dimension-full minimal scale factors in
	noncomutative phase space, which help us to maintain consistent  dimension of all
	SU(1,1) generators in (\ref{alg}) and  in terms of the dimentionless basis
	$\boldsymbol{T}_{\mu} $ ( where $\mu=0,1,2 $), the above (\ref{algcom})
	commutation relations take a more suggestive form as,
	
	\begin{equation}
	[\boldsymbol{T_{0}},\boldsymbol{T_{i}}]=i \tilde{\hbar}\epsilon_{ij} 
	\boldsymbol{T_{j}};~~~~~ [\boldsymbol{T_{i}},\boldsymbol{T_{j}}]=-i\tilde{\hbar}
	\epsilon_{ij}\boldsymbol{T_{0}}~~~~(i,j=1,2),
	\end{equation}
	where $\tilde{\hbar}= \frac{\hbar}{\sqrt{-\theta\eta}}$ is the dimentionless reduced
	Planck's constant. Notice at this stage that $\boldsymbol{T_{0}}$ and
	$\boldsymbol{T_{1}}$ are skew hermitian like $\boldsymbol{K_{\pm}}$ and
	$\boldsymbol{K_{0}}$, but $\boldsymbol{T_{2}}$ is hermitian.\\\\
	A faithful finite 2D representation
	\cite{doi:10.1063/1.528100,doi:10.1119/1.17149}  "$\Pi$" of this SU(1,1) is
	furnished by the Pauli matrices $\vec{\boldsymbol{\sigma}}$'s 
	as\footnote{Observe at this stage that in this finite dimensional representation, it is rather $\Pi({\boldsymbol{T}_{0}})$ which is only hermitian and $\Pi({\boldsymbol{T}_{i}})$'s are skew-hermitian. This is a typical and peculiar feature of finite-dimesional representations of the Lie-algebra associated with non-compact unitary groups like SU(1,1).}
	
	\begin{equation}
	\Pi({\boldsymbol{T}_{0}})=\frac{\tilde{\hbar}}{2}
	\sigma_{3},\Pi({\boldsymbol{T}_{i}})=- \frac{i\tilde{\hbar}}{2}\sigma_{i}
	\end{equation}
	Using this representation, one can easily see that, any su(1,1) Lie-algebra
	element $ A^{\mu}  \Pi(\boldsymbol{T}_{\mu})$, with coefficients $A^\mu$ and
	$\mu=0,1,2$, takes the following trace-less form
	\begin{equation}
	A^{\mu}  \Pi(\boldsymbol{T}_{\mu})=\frac{\tilde{\hbar}}{2}\begin{pmatrix}
	A^{0}& -A \\
	A^{*} & -A^{0} 
	\end{pmatrix} ;~~~~A=A^{1}+iA^{2}
	\end{equation}
	
	If this object is now subjected to a SU(1,1) transformation  by $\mathcal {U}
	\in SU(1,1)$ as
	\begin{equation}
	A^{\mu}  \Pi(\boldsymbol{T}_{\mu})\rightarrow \mathcal {U}A^{\mu} 
	\Pi(\boldsymbol{T}_{\mu})\mathcal {U}^{\dagger}:=
	B^{\mu}\Pi(\boldsymbol{T}_{\mu}), 
	\label{Amu}
	\end{equation}\\
	(where we could easily replace $\mathcal {U}\rightarrow \mathcal{V}\in
	SU(1,1)\otimes U(1) $, as $[\boldsymbol{L},\boldsymbol{K_{\mu}}]=0~~ \forall
	~\mu$.) the resulting object in (\ref{Amu}) will again be another $su(1,1) $ element with some other coefficeient $B^{\mu}$, where the trace-less ness property will be preseved along with the determinant.
	Particularly, the latter i.e the preservation of determinant implies that we must have the following identity holding:
	\begin{equation}
	(A^{1})^{2}+(A^{2})^{2}-(A^{0})^{2}=(B^{1})^{2}+(B^{2})^{2}-(B^{0})^{2}.
	\label{sop}
	\end{equation}
	We immediately conclude that $A^{\mu}$ can be regarded as a 3-vector transforming under Lorentz
	transformation SO(2,1) in (2+1)D with $A^{i}$'s and $A^{0}$ may be thought of representing
	spatial and temporal components respectively. This connection of the Lorentz group $SO(2,1)$ with its double cover $SU(1,1)$ or $SL(2,\mathbb{R})$ is well known in the literature: $SO(2,1)=SU(1,1)/\mathbb{Z}_2=SL(2,\mathbb{R})/\mathbb{Z}_2$; all of them share the same Lie algebra.\\
	
	We can now apply all of these formalisms to our system Hamiltonian (\ref{comh})
	which can be written as a linear combination of the original infinite-dimensional
	representation of $SU(1,1)\otimes U(1)$ group generators (\ref{alg}) as,
	
	\begin{equation}
	\mathcal{H}(t)=-2i[\alpha(t)\boldsymbol{K_{-}} +\beta(t)
	\boldsymbol{K_{+}}+2\delta(t)\boldsymbol{K_{0}}]-\gamma(t)\boldsymbol{L}=\mathcal{H}_{gho}(t)-\gamma(t)\boldsymbol{L}
	\label{gro}
	\end{equation}
	
	Clearly the part of Hamiltonian $H_{gho}(t)$ in (\ref{hghoi})
	is an $su(1,1)$ Lie algebra valued element. Re-writing this in 		    	terms of generators 
	$\boldsymbol{T}_{\mu}:=(\boldsymbol{T}_{0},\boldsymbol{T}_{1},\boldsymbol{T}_{2})$
	introduced in (\ref{dimalg}) we get
	
	\begin{equation}
	\mathcal{H}_{gho}(t)=-2i A^{\mu}(t)\boldsymbol{T}_{\mu}
	\label{ha},
	\end{equation}\\
	
	where

	\begin{equation}
	A^{\mu}(t)=\begin{pmatrix}
	A^{1}(t) \\
	A^{2}(t) \\
	
	A^{0}(t) \\
	\end{pmatrix}=\begin{pmatrix}
	(-\eta\alpha(t)+\theta\beta(t)) \\
	i(\eta\alpha(t)+\theta\beta(t)) \\
	(2\sqrt{-\theta\eta}\delta(t)) \\
	\end{pmatrix}
	\label{Atr}
	\end{equation}
	Note that $A^2(t)$, as occurs here is purely imaginary and this 	ensures the Hermiticity of $\mathcal{H}_{gho}(t)$ (\ref{ha}).
	Now if the above 3-vector $A^{\mu}(t)$  scaled by $\tilde{\hbar}$ as
	\begin{equation}
	A^{\mu}(t)\rightarrow \tilde{A}^{\mu}(t):=\tilde{\hbar}A^{\mu}(t)
	\end{equation}

	invariance of SO(2,1) norm of $\tilde{A}^{\mu}(t)$ readily gives
	\begin{equation}
	(\tilde{A}^{1}(t))^{2}+(\tilde{A}^{2}(t))^{2}-(\tilde{A}^{0}(t))^{2}=4\hbar^{2}[\alpha(t)\beta(t)-\delta^{2}(t)]=\hbar^{2}
	\omega^{2}(t)>0,
	\label{su}
	\end{equation}
	
	where $\omega(t)>0$ in (\ref{ttt}) is the frequency of the $\mathcal{H}_{gho}(t)$ and is invariant under the instantaneous Lorentz transformation.\\
	
	Further $2\delta(t)\propto A^0(t)$ here in (\ref{Atr},\ref{su})  can be identified with the  temporal component of the space-like 3-vector
	$A_{\mu}$. Consequently, at any instant
	$t=t_{0}$, the tip of the 3-vector $A^{\mu}(t_{0})$ will lie on a 2D-hyperboloid
	whose tangent plane is orthogonal to $A^{\mu} $ and as time evolves $A^{\mu}$
	will trace out
	a trajectory (in fact a closed loop here) intersecting this one-parameter family of such hyperboloids.
	Further, the space-like nature of $A^{\mu}$ implies that at the instant
	$t=t_{0}$ we can choose a suitable SO(2,1) transformation such that $\delta(t_0)$
	vanishes in this particular Lorentz frame. Obviously this needs to change from
	moment to moment and therefore be time-dependent. This, on the other hand, can
	be induced by subjecting the Hamiltonian
	$\mathcal{H}_{gho}(t)$(\ref{gro},\ref{ha}) a time dependent unitary
	transformation belonging to the covering group $\mathcal{W}(t)\in SU(1,1)$ in the manner of
	(\ref{Amu}). One can, of course, consider the bigger group $SU(1,1)\otimes U(1)$ also here, but the presence of U(1) is quite inconsequential here and therefore optional in nature. This has to be contrasted with (\ref{funda}) in section-4, where we need to choose a specific U(1) element other than identity element.
	
	In fact it is not very difficult to construct such a unitary operator 
	$\mathcal{W}(t)$. To that end consider an instantaneous SO(2,1)
	transformation $\Lambda (t_{0})$ transforming the triplet
	\begin{equation}
	(\alpha(t_{0}),\beta(t_{0}),\delta(t_{0}))\rightarrow(\alpha^{'}(t_{0}),\beta^{'}(t_{0}),\delta^{'}(t_{0})):=(\alpha(t_{0}),\beta^{'}(t_{0}),0)
	\label{cotra}
	\end{equation}
	in such a manner that the coefficient of the dilatation term vanishes. Using
	(\ref{su}), we readily obtain
	\begin{equation}
	\beta^{'}(t_{0})=\frac{\alpha(t_{0})\beta(t_{0})-\delta^{2}(t_{0})}{\alpha^{'}(t_{0})}
	\end{equation}
	
	One can easily verify, at this stage, that a corresponding unitary
	transformation $\mathcal{W}(t)$ at an arbitary time $t$ can be constructed as,
	\begin{equation}
	\mathcal{W}(t)=\exp[{\frac{i}{\hbar}\frac{\delta(t)}{2\alpha(t)}\vec{q}^{2}}]
	\label{W}
	\end{equation}
	Under this transformation the instantaneous total Hamiltonian $ \mathcal{H}(t)$
	(\ref{hsep}) indeed transforms as 
	\begin{equation}
	\mathcal{H}(t)\rightarrow\mathcal{W}(t)\mathcal{H}(t)\mathcal{W}^{\dagger}(t)=\alpha(t)p^{2}_{i}+(\frac{\alpha(t)\beta(t)-\delta^{2}(t)}{\alpha(t)})
	q^{2}_{i}-\gamma(t)\epsilon_{ij}q_{i}p_{j}
	\label{htra}
	\end{equation}
	eliminating the dilatation term. However, from the time dependent Schrodinger equation,
	one can
	easily recognize that (\ref{htra}) should not be identified as the Hamiltonian
	responsible for
	the time evolution of the transformed instantaneous state
	($\mathcal{W}(t)\left|\Psi(t)\right\rangle $) as $\mathcal{W}(t)$ itself has an
	explicit time dependence. Indeed, from it is not difficult to see that the time
	evolution of the state ($\mathcal{W}(t)\left|\Psi(t)\right\rangle $) is governed by the effective Hamiltonian
	$\tilde{\mathcal{H}}(t)$, obtained by augmenting the one in (\ref{htra}) by a
	suitable "connection" term as,
	\begin{equation}
	\mathcal{H}(t)\rightarrow\tilde{\mathcal{H}}(t)=\mathcal{W}(t)\mathcal{H}(t)\mathcal{W}^{\dagger}(t)-i\hbar\mathcal{W}(t)\frac{d}{dt}\mathcal{W}^{\dagger}(t)
	\label{connection}
	\end{equation}
	so that $i\hslash \partial_{t}(W(t)\left|\psi(t)\right \rangle)=\tilde{\mathcal{H}}(t)(W(t)\left|\psi(t)\right \rangle)$ holds. This has to be contrasted with the case involving time-independent unitary transformation $\boldsymbol{U}$ (\ref{unitary}) (See discussion below (\ref{SUP})) connecting $ \mathcal{H}^{(1)}(t)$ and $\mathcal{H}^{(2)}(t)$ (\ref{funda}). Expanding this $\tilde{\mathcal{H}}(t)$ we obtain
	
	\begin{equation}
	\tilde{\mathcal{H}}(t)=\alpha(t)p^{2}_{i}+(\frac{\alpha(t)\beta(t)-\delta^{2}(t)-\frac{\alpha}{2}\frac{d}{dt}(\frac{\delta(t)}{\alpha(t)})}{\alpha(t)})q^{2}_{i}-\gamma(t)\boldsymbol{L}=\mathcal{H}_{sho}(t)-\gamma(t)\boldsymbol{L}
	\label{d}
	\end{equation}
	
	This is like a usual 2D harmonic oscilator Hamiltonian with just the Zeeman  coupling
	$\gamma(t)\boldsymbol{L}$. Here we also observe that 
	$\gamma(t)\boldsymbol{L}$, $\mathcal{H}_{sho}(t)$ and $\tilde{\mathcal{H}}(t)$ commutes
	among each other, even at different times.
	So, they have simultaneous instantaneous eigenstates.
	
	To find out the eigenstates of the system Hamiltonian (\ref{d}) one may proceed and
	introduce the annihilation (and corresponding creation) operator. This can just be obtained by setting $\delta=0$ and replacing $\beta\rightarrow\tilde{\beta}$ in (\ref{ol}) to get,
	\begin{equation}
	\tilde{\boldsymbol{a}}_{j}=\left(\frac{\tilde{\beta}}{4\alpha\hbar^{2}}\right)^{\frac{1}{4}} \left[
	q_{j}+i\sqrt{\frac{\alpha}{\tilde{\beta}}}p_{j}\right];j=1,2
	\end{equation}
	with
	$\tilde{\beta}=(\frac{\alpha(t)\beta(t)-\delta^{2}(t)-\frac{\alpha}{2}\frac{d}{dt}(\frac{\delta(t)}{\alpha(t)})}{\alpha(t)})$,
	satisfying the commutation relation
	$[\hat{\tilde{\boldsymbol{a}}}_{j},\hat{\tilde{\boldsymbol{a}}}_{k}^{\dagger}]=\delta_{jk}$
	Accordingly, the system Hamiltonian (\ref{d}) can be written as
	\begin{equation}
	\tilde{\mathcal{H}}(t)=\hbar\tilde{\omega}(t)(\tilde{\boldsymbol{a}}^{\dagger}_{j}\tilde{\boldsymbol{a}}_{j}+1)+i\hbar\gamma(t)\epsilon_{jk}\tilde{\boldsymbol{a}}^{\dagger}_{j}\tilde{\boldsymbol{a}}_{k}~~~~(j,k)\in \{1,2\}
	\label{a}
	\end{equation}
	where
	\begin{equation}
	\tilde{\omega}(t)=2\sqrt{(\alpha(t)\beta(t)-\delta^{2}(t))-\frac{\alpha(t)}{2}\frac{d}{dt}(\frac{\delta(t)}{\alpha(t)})}.
	\label{arxiv}
	\end{equation}
	Again introducing operators $\boldsymbol{a}_{\pm}$ through time
	independent canonical transformation (\ref{eq:J2 to J3}) we get  the
	Hamiltonian (\ref{a})   in  standard diagonal Hermitian form as,
	\begin{equation}
	\tilde{\mathcal{H}}=\hbar\tilde{\omega}(\tilde{\boldsymbol{a}}^{\dagger}_{j}\tilde{\boldsymbol{a}}_{j}+1)-\hbar\gamma(t)(\tilde{\boldsymbol{a}}^{\dagger}_{+}\tilde{\boldsymbol{a}}_{+}-\tilde{\boldsymbol{a}}^{\dagger}_{-}\tilde{\boldsymbol{a}}_{-}); ~~~~~~~~~j\in \{+,-\}.
	\end{equation}
	
	 The instantaneous nondegenerate eigenstates are now given by,
	\begin{equation}
	\begin{aligned}
	&~~~~~~~~~~~~\left|n_{+},n_{-};(t)\right\rangle
	=\frac{\left(\tilde{\boldsymbol{a}}^{\dagger}_{+}\right)^{n_{+}}\left(\tilde{\boldsymbol{a}}^{\dagger}_{-}\right)^{n_{-}}}{\sqrt{n_{+}!}\sqrt{n_{-}!}}\left|0,0;(t)\right\rangle;~~\tilde{\boldsymbol{a}}_{\boldsymbol{\pm}}(t)\left|0,0;(t)\right\rangle =0.\\
	\label{sp}
	\end{aligned}
	\end{equation}
	Note that we have written the time argument $t$ within the parenthesis as $(t)$ in order to distinguish these tower of eigen states from (\ref{ene}) which are clearly not the same; they are build upon different instantaneous ground states.
	
	Correspondingly the instantaneous discrete  energy-eigenvalues are given by,
	\begin{equation}
	\begin{aligned}
	E_{n_{+},n_{-}}(t) &
	=\hbar\tilde{\omega} (n_{+}+n_{-}+1)-\hbar\gamma(t)(n_{+}-n_{-})\\
	&
	\simeq\hbar(n_{+}+n_{-}+1)\omega(t)\left[1-\frac{\alpha(t)}{\omega(t)^{2}}\frac{d}{dt}(\frac{\delta(t)}{\alpha(t)})\right]-\hbar\gamma(t)(n_{+}-n_{-})
	\end{aligned}
	\label{c}
	\end{equation}
	where $\omega=2\sqrt{\alpha\beta-\delta^{2}}$ . Here it suffices to work in the
	first order of adiabaticity (manifested through the order of time derivatives of the
	parameters). This is tantamount to ignoring the higher order time derivatives of
	the slowly varying parameters in (\ref{arxiv},\ref{c}).\\
	
	Now since $\tilde{\mathcal{H}}(t)$ and $\mathcal{H}(t)_{sho}$ in (\ref{d}) commutes with each other, they share the same eigenspaces.
	Consequently, we can express an eigenstate (\ref{sp})  of $\tilde{\mathcal{H}}(t)$, as
	a linear of combination of eigenstates of $\mathcal{H}(t)_{sho}$, as,
	\begin{equation}
	\left|n_{+},n_{-};(t)\right\rangle=\sum_{n_1+n_2=n_{+}+n_{-}}
	C_{n_1,n_2}^{n_{+},n_{-}} \left|n_1,n_2;(t)\right\rangle_{sho}^{2D}
	\end{equation}

	where we have denoted the eigenstates of $\mathcal{H}(t)_{sho}$ as
	\begin{equation}
	\left|n_1,n_2;(t)\right\rangle_{sho}^{2D}=\frac{\left(\tilde{\boldsymbol{a}}^{\dagger}_{1}\right)^{n_{1}}\left(\tilde{\boldsymbol{a}}^{\dagger}_{2}\right)^{n_{2}}}{\sqrt{n_{1}!}\sqrt{n_{2}!}}\left|0,0;(t)\right\rangle
	\end{equation}

	and $n_1+n_2=n_{+}+n_{-}$. This restriction ensures that the eigenstates to be taken from a single
	eigenspace of $\mathcal{H}(t)_{sho}$ \cite{book:17486}. Note that, $C_{n_1,n_2}^{n_{+},n_{-}}$'s
	are themselves time independent, because the ladder operators diagonalising
	$\tilde{\mathcal{H}}(t)$ are derivable from the ladder operators of
	$\mathcal{H}(t)_{sho}$, using time-indepedent invertible linear transformation
	(\ref{eq:J2 to J3}). This also ensures that, both sets of lowering operators $\{\tilde{\boldsymbol{a}}_{1},\tilde{\boldsymbol{a}}_{2}\}$ or ${\tilde{\boldsymbol{a}}_{\pm}}$ annihilates the
	same instantaneous ground state $\left |0,0;(t)\right\rangle$ at time $t$.\\
	Since we are working in the regime where adiabatic theorem works properly, the
	Berry phase for an eigenstate $\left|n_{+},n_{-}\right\rangle$, if exists, would
	be given by:
	
	\begin{equation}
	\begin{aligned}
	\Phi^{(G)}_{(n_{+},n_{-})} &=-i \int d t\left\langle n_{+},n_{-};(t)\left|\frac{d}{d
		t}\right| n_{+},n_{-};(t)\right\rangle\\
	&=-i \int d t \sum_{\substack{n_1+n_2=m_1+m_2 \\ 
=n_{+}+n_{-}
	}}C_{m_1,m_2}^{n_{+},n_{-}\star}C_{n_1,n_2}^{n_{+}n_{-}}~
	{^{2D}_{sho}\left\langle m_1,m_2(t)\right|}\frac{d}{d t}\left|
	n_1,n_2;(t)\right\rangle^{2D}_{sho}
	\end{aligned}
	\end{equation}
	
	Now for a pair of tuples $(m_1,m_2)$ and $(n_1,n_2)$ sandwiching inside the
	sum, there are two possibilities: either (i) $m_1=n_1, m_2=n_2$  or (ii) $m_1\neq n_1,
	m_2\neq n_2$. 
	
	Let us consider the case of second possibility (ii) first as, 
	\begin{equation}
	\begin{aligned}
	{^{2D}_{sho}\left\langle m_1,m_2;(t)\right|}\frac{d}{d t}\left|
	n_1,n_2;(t)\right\rangle^{2D}_{sho}
	=\left({^{1D}_{sho}\left\langle m_1(t)\right|}\frac{d}{d t}\left|
	n_1(t)\right\rangle^{1D}_{sho}\right) \times
	\left({^{1D}_{sho}}\left\langle m_2(t)|n_2(t)\right\rangle_{sho}^{1D}\right)\\
	+\left({^{1D}_{sho}\left\langle m_2(t)\right|}\frac{d}{d t}\left|
	n_2(t)\right\rangle^{1D}_{sho}\right) \times
	\left({^{1D}_{sho}}\left\langle m_1(t)|n_1(t)\right\rangle_{sho}^{1D}\right) = 0
	\end{aligned}
	\end{equation}
	
	Whereas for the case of first possibility (i) $(m_1,m_2)=(n_1,n_2)$, we get,
	\begin{equation}
	\begin{aligned}
	\int dt \left({^{2D}_{sho}\left\langle n_1,n_2;(t)\right|}\frac{d}{d t}\left|
	n_1,n_2;(t)\right\rangle^{2D}_{sho} \right) &=
	\int dt \left({^{1D}_{sho}\left\langle n_1(t)\right|}\frac{d}{d t}\left|
	n_1(t)\right\rangle^{1D}_{sho}\right.\\
    &\left. +{^{1D}_{sho}\left\langle n_2(t)\right|}\frac{d}{d
		t}\left| n_2(t)\right\rangle^{1D}_{sho}\right)=0.
	\end{aligned}
	\end{equation}
	
	as this represents the Berry phase of a pair of de-coupled 1D simple harmonic oscillators which we know to have vanishing Berry phase. So $\Phi^{(G)}_n=0$, implying the total
	hamiltonian $\tilde{\mathcal{H}}(t)$  does not produce any Berry phase by itself, apparently.\\
	
	However, the total dynamical phase acquired by
	$\left|n_{+},n_{-};(t)\right\rangle$ after a complete cycle $\Gamma$ of time period $T$ by the Hamitonian $\tilde{\mathcal{H}}(t)$, is obtained by using (\ref{c}), to get: 
	
	\begin{equation}
	\begin{aligned}
	\Phi_{n_{+},n{-}}(T) &=\int_{0}^{T} dt  \frac{ E_{n_{+},n_{-}}(t)}{\hbar}\\
	&= \int_{0}^{T} dt
	\left[(n_{+}+\frac{1}{2})(\omega+\gamma)+(n_{-}+\frac{1}{2})(\omega-\gamma)-(n_{+}+n_{-}+1)\frac{\alpha}{\omega}\frac{d}{dt}\left(\frac{\delta}{\alpha}\right)\right],
	\end{aligned}
	\end{equation}
	
	%
	%
	%
	
	
	agreeing with our result (\ref{eq:adiaphase}) after a gauge transformation
	(\ref{phg}). Indeed the last term, although now occurs in the dynamical phase,
	it nevertheless retains its geometric character as it represents line integral of the one-form $\boldsymbol{A}$ (\ref{phg}) along the closed loop $\Gamma$ as $\int_{\Gamma}\boldsymbol{A}$ and therefore the resulting phase is a functional of $\Gamma$ and matches exactly with the Berry phase (\ref{fp}).
	This whole excercise therefore shows how one can eliminate the crucial dilatation term responsible for the Berry phase through a time-dependent unitary
	trasnformation to find it to reappear again in "disguise" within the dynamical part,
	revealing its geometric origin, when considering the total adiabatic phase as
	a whole. It should therefore show up in suitably designed interference experiments.

\end{large}

 \bibliographystyle{ieeetr}
\bibliography{bib}

\end{document}